\documentclass[pdflatex,sn-mathphys-num]{sn-jnl}% Math and 

\usepackage{graphicx}%
\usepackage{multirow}%
\usepackage{amsmath,amssymb,amsfonts}%
\usepackage{amsthm}%
\usepackage{mathrsfs}%
\usepackage[title]{appendix}%
\usepackage{xcolor}%
\usepackage{textcomp}%
\usepackage{manyfoot}%
\usepackage{booktabs}%
\theoremstyle{thmstyleone}%
\newtheorem{theorem}{Theorem}%  meant for continuous numbers
\theoremstyle{thmstyletwo}%
\newtheorem{example}{Example}%

\theoremstyle{thmstylethree}%
\newtheorem{definition}{Definition}%

\usepackage{listings}
\usepackage{lscape}
\usepackage[ruled,vlined,linesnumbered]{algorithm2e}

\newcommand{\Deg}{{\it deg}} %%%%% differ from \deg that is used. 
\newcommand{\core}{{\it core}}
\usepackage{enumitem}
    \setlist[itemize]{label=--} % Dash as bullet

\begin{document}

\title[Experimental Evaluation of Distributed $k$-Core Decomposition]{Experimental Evaluation of Distributed $k$-Core Decomposition}
% \author{
% Bin Guo \\
% Department of Computer Science, Trent University 

% Runze Zhao \\
% {Department of Computing \& Software}, xtit{McMaster University}
% }
%\date{November 2024}

% \author{Bin Guo, \url{binguo@trentu.ca},
% Department of Computer Science, Trent University\\

% Runze Zhao, Department of Computing \& Software, McMaster University

% }


\author*[1]{\fnm{Bin} \sur{Guo}}\email{binguo@trentu.ca}
\equalcont{These authors contributed equally to this work.}

% \author[2]{\fnm{Emil} \sur{Sekerinski}}\email{emil@mcmaster.ca}

\author[2]{\fnm{Runze} \sur{Zhao}}\email{zhaorz@mcmaster.ca}
\equalcont{These authors contributed equally to this work.}

\affil*[1]{\orgdiv{Department of Computer Science}, \orgname{Trent University}, \orgaddress{\street{1600 W Bank Dr}, \city{Peterborough}, \postcode{K9L 0G2}, \state{ON}, \country{Canada}}}

\affil[2]{\orgdiv{Department of Computer and Software}, \orgname{McMaster University}, \orgaddress{\street{1280 Main St W}, \city{Hamilton}, \postcode{L8S 4L8}, \state{ON}, \country{Canada}}}

\abstract{
Given an undirected graph, the $k$-core is a subgraph in which each node has at least $k$ connections. This is widely used in graph analytics to identify core subgraphs within a larger graph. The sequential $k$-core decomposition algorithm faces limitations due to memory constraints, and many data graphs are inherently distributed. A distributed approach is proposed to overcome limitations by allowing each vertex to compute its core number independently using only local information. This work explores the experimental evaluation of a distributed $k$-core decomposition algorithm. By assuming that each vertex is a client as a single computing unit, we simulate the process using Golang, leveraging its Goroutines and message passing. Since real-world data graphs can be large with millions of vertices, it is expensive to build a distributed environment with millions of clients if experiments were run in a real distributed environment. Therefore, our experimental simulation can effectively evaluate the running time and message passing for the distributed $k$-core decomposition. 

% This study implements and simulates the distributed k-core algorithm using Golang, leveraging its concurrency features to handle large-scale graphs efficiently. Experiments were conducted on graphs with varying sizes, from thousands to millions of vertices, to evaluate the runtime behavior, message complexity, and overall performance of the distributed algorithm. The findings demonstrate the viability of distributed k-core decomposition for large graphs, highlighting its scalability and efficiency compared to sequential approaches.
}


\maketitle

% \author[1]{Bin Guo}{first.author@example.com}}}
% \author[2]{Second Author\thanks{\href{mailto:second.author@example.com}{second.author@example.com}}}
% \affil[1]{Department of Studies, University Z}
% \affil[2]{Research Institute, City}

\section{Introduction}

Graphs are fundamental data structures that are used to model real-world applications. They represent mathematical relationships between objects. In a graph, each vertex represents an object, and each edge represents a relationship between a pair of objects. For example, social networks can be represented as graphs, where each user is represented by a vertex, and the relationships between users are represented by edges.
Since many real-world applications can be modeled as graphs, graph analytics has attracted considerable attention from both research and industry communities. Several algorithms have been proposed to analyze large data graphs, including graph trimming, strong connected component (SCC) decomposition, $k$-core decomposition, and $k$-truss decomposition. 

Among all the above algorithms, the \emph{$k$-core decomposition} analyzes the graph structure by identifying its core subgraphs. The $k$-core of a graph is a subgraph in which each node has the least $k$ connections to other nodes; the core number of each vertex is the highest $k$ of a subgraph in which each vertex has at least $k$ neighbors.
There are various applications~\cite{Kong2019} for the $k$-core decomposition:
\begin{itemize}

    \item \emph{Biology.} The $k$-core and phylogenetic analysis of the Protein-Protein Interaction network helps predict the characteristics of unknown functional proteins.
    \item \emph{Social Networks.} The $k$-core decomposition is widely accepted as revealing the structure of the network. It can be used to identify the key nodes in the network or to measure the influence of users on online social networks.
    \item \emph{Computer Sciences}. The $k$-core decomposition can be used to study large-scale Internet graphs. It easily reveals the ordered and structural features of the networks. The ~$k$-core subgraphs also reveal the primary hierarchical layers of the network and allow their analytical characterization.
\end{itemize}

%\textbf{[simplify this discussion, remove the details, as we can do the details in the Section 2. ]}

The widely used algorithm for the $k$-core decomposition was proposed by Batagelj and Zaversnik, the so-called BZ algorithm~\cite{bz2003}. It recursively removes vertices (and incident edges) with degrees less than $k$. The algorithm uses bucket sorting and can run in $O(m +n)$ time, where $m$ is the number of edges and $n$ is the number of vertices. 
The sequential BZ algorithm has two limitations:
First, the entire graph must be loaded into memory, as the BZ algorithm requires random access to the entire graph during computation. Due to memory restrictions, some graphs may be too large to fit in a single host. 
Second, the graph can be inherently distributed over a collection of hosts. In which case, it is not convenient to move each portion to a central host. 
To overcome the above limitations, distributed graph algorithms are proposed in~\cite{montresordistributed}. Specifically, each vertex in the graph is a standalone host that calculates its own $k$-core simultaneously, which means that $k$-core decomposition happens in a distributed environment. Each vertex does not need to store the information of the entire graph; instead, it only stores the core number of its neighbours. Hence, less memory is required for each vertex. In addition, each vertex only calculates its own core number. Compared to the centralized algorithm, where all calculations take place in a single host, each vertex now performs fewer calculations, resulting in less computing power required on each vertex.

%%%% from chatgpt about motivation 
Although distributed approaches address scalability, evaluating them on real infrastructures poses major challenges: deploying millions of computational units across a physical cluster is costly, logistically complex, and often impossible in research settings. Moreover, such deployments make it difficult to isolate and study the precise effects of algorithmic design choices, such as message complexity or termination detection.

To overcome these barriers, this work simulates the distributed $k$-core decomposition algorithm. By leveraging Golang's lightweight concurrency (Goroutines) and message-passing mechanisms, we can realistically mimic the behaviour of a fully distributed system within a single machine. This allows us to systematically evaluate scalability, message complexity, and runtime characteristics on real-world graphs, providing insights that would otherwise require prohibitively expensive distributed infrastructure. 
Specifically, our experiment is conducted with Golang to simulate a distributed run-time environment. We utilize real-world data graphs, with sizes ranging from thousands to millions of vertices. The core number of each vertex is calculated, and we capture the messages passed between each vertex for analysis.
We summarize our contribution as follows:
\begin{itemize}
    \item We review the existing distributed $k$-core decomposition algorithm, by adding detailed explanations of message passing (Section~\ref{Review}).
    \item We implement the algorithm using Golang for simulating real distributed environments (Section~\ref{implementation}).
    \item We evaluate the algorithm with real data graphs, including the number of messages, the number of active nodes, running time, and scalability (Section~\ref{experiment}). 
\end{itemize}

%%%% about the motivation. 

\iffalse 
In this work, we simulate the distributed $k$-core decomposition algorithm to explore its runtime behaviours. The experiment is conducted with Golang to simulate a distributed run-time environment. In our experiment, we used real-world data graphs. The size of data graphs varies from thousands to millions of vertices. The core number of each vertex is calculated, and we capture the messages passed between each vertex for analysis. 
\fi 

This work is structured as follows. 
The distributed $k$-core decomposition algorithm is explained in Section \ref{preliminaries}. We review the existing $k$-core decomposition algorithm in Section~\ref{Review}. The details of the implementation are demonstrated in Section \ref{implementation}. Our experiments are described in Section \ref{experiment}. The conclusion and future work are summarized in Section~\ref{conclusion}.

\section{Preliminaries} \label{preliminaries}

In this section, we review the distributed $k$-core decomposition algorithm.
Let $G = (V, E)$ be an undirected unweighted graph, where $V(G)$ denotes the set of vertices and $E(G)$ represents the set of edges in $G$. When the context is clear, we will use $V$ and $E$ instead of $V(G)$ and $E(G)$ for simplicity, respectively.
As $G$ is an undirected graph, an edge $(u, v)\in E(G)$ is equivalent to $(v, u)\in E(G)$. 
We denote the number of vertices and edges of $G$ by $n$ and $m$, respectively. 
The set of neighbours of a vertex $u \in V$ is defined by $u.adj = \{v \in V: (u, v) \in E\}$.
The degree of a vertex $u\in V$ is denoted by $u.\Deg = |u.adj|$.

\subsection{The $k$-Core Decomposition}

\begin{definition} [$k$-Core] \label{def: k-core}
Given an undirected graph $G=(V, E)$ and a natural number $k$, an induced subgraph $G_k$ of $G$ is called a $k$-core if it satisfies: (1) all vertices in $G_k$ have degrees not smaller than $k$ defined as $\forall u \in V(G_k): u.\Deg \geq k$, and (2) $G_k$ is a maximum subgraph defined as $\nexists u \in V\setminus V(G_k)$ such that $\forall v\in V(G_k)\cup \{u\}: v.\Deg \geq k$. 
%%% here is G_k is a maximum subgraph. 
% can be defined not exist a u that canbe added to G_k and satisify (1)

%(2) for all natural numbers $k$ that satisfy the previous condition, the $G_k$ that has maximum $k$ is the $k$-core of $G$. Moreover, $G_{k+1} \subseteq G_k$, for all $k \geq 0$, and $G_0$ is just $G$.
\end{definition}

\begin{definition}[Core Number] \label{def: core number}
Given an undirected graph $G=(V,E)$, the core number of a vertex $u\in G(V)$, denoted $u.\core$, is the largest value of $k$ for $u\in G_k$, defined as $u.\core = k: u\in G_k \land u\notin G_{k+1}$. 
\end{definition}

%In other words, $u.\core$ is the largest $k$ so that there exists a $k$-core containing $u$.

\begin{definition} [$k$-Core Decomposition] \label{def: k-core decom}
Given a graph $G=(V,E)$, the problem of computing the core number for each $u \in V(G)$ is called $k$-core decomposition. 
\end{definition}

\begin{figure}[!htb]
    \centering
    \includegraphics[width=0.6\linewidth]{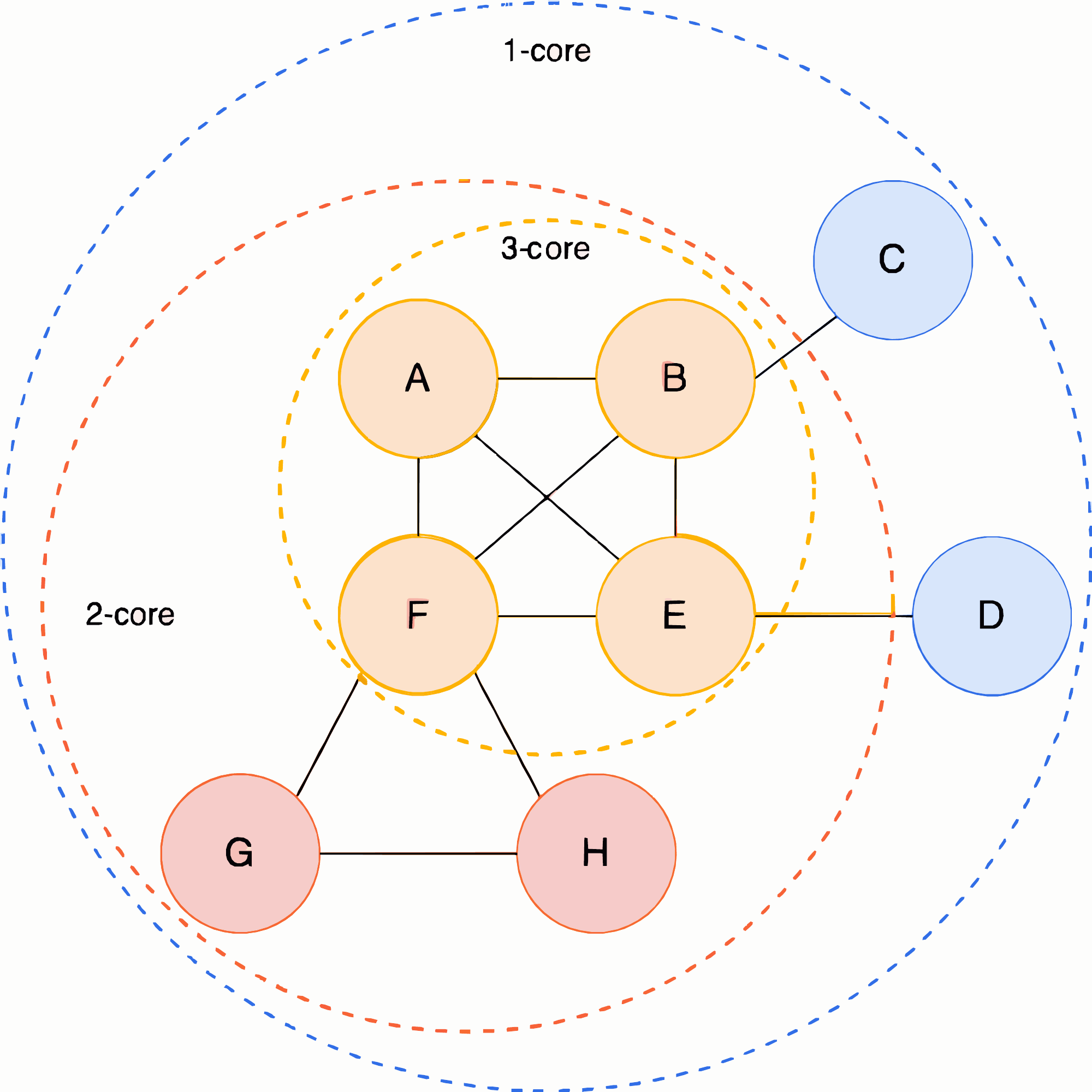}
    \caption{$k$-Core Decomposition}
    \label{fig:K-core decomposition}
\end{figure}

\begin{example}[$k$-core decomposition]
Figure~\ref{fig:K-core decomposition} demonstrates the basic $k$-core decomposition. The coloured circles $A, B, C, D, E, F, G, H$ are the graph nodes. The black lines that connect two nodes are the edges. The edges are undirected. The node $A$ connects with 3 edges, which means that it has a degree of 3. 

The three dashed circles demonstrate the $k$-cores of this graph. As mentioned, the $k$-core is a maximal subgraph in which every node has at least degree $k$. For example, nodes $A, B, E, F, G, H$ form the subgraph where each node has at least a degree of 2. Hence, this subgraph is $2$-core. Similarly, the subgraph containing nodes $A, B, E, F$ is $3$-core because each node in this subgraph has at least a degree of 3. 

The core number of a node is the largest $k$ such that this node is part of the $k$-core. In our example, node $A$ is part of $1$-core, $2$-core and $3$-core; the largest $k$ is 3, hence the core number of node $A$ is 3. Similarly, nodes $B, E, F$ have a core number of 3; nodes $G, H$ have a core number of 2 and nodes $C, D$ have a core number of 1.
\end{example}

\subsection{Algorithm Notations}
In this work, all algorithms use the \emph{Communicating Sequential Process} (CSP)~\cite{10.1145/359576.359585} notation. 
Table~\ref{csp annotation} explains the functions of different operators in the algorithm.
\begin{table}[!htb]
    \centering
    \begin{tabular}{|c|c|} \hline 
         Syntax& Function\\ \hline 
         $c!v$&send the value v on the channel c\\ \hline 
         $c?x$&Receive x from channel c\\ \hline 
         $a \to P$& perform the event a, then continue as the process P\\ \hline 
         $a \gets v$& assign value v to variable a\\ \hline
         $a \nrightarrow b$ & a map, in which a is the key and b is the value \\ \hline
    \end{tabular}
    \caption{CSP Annotation}
    \label{csp annotation}
\end{table}

\section{Review Existing Approach}
\label{Review}
In this section, we review the existing distributed $k$-core decomposition algorithm in~\cite{montresordistributed}. We rewrite the algorithm with the CSP notation, which fully describes the details of message passing. 
%as  is based on the existing distributed $k$-core decomposition approach .

\begin{theorem}[Locality~\cite{montresordistributed}]
For all $u\in V$, $u.core$, as in Definition \ref{def: core number} then $u.core = k$, where
$k \leq   |\{v\in u.adj: v.core \geq k\}|$ and $ (k+1) > |\{v\in u.adj: v.core \geq (k+1)\}| $.
\label{th:locality}
\end{theorem}

Theorem \ref{th:locality} shows that the core number of each vertex $u$ can be calculated by comparing its core number with the core numbers of its neighbours.

\paragraph{Algorithm.}
The procedure works as in the following steps, and the details are shown in Algorithm~\ref{alg:go core-decompotion}.
\begin{enumerate}
    \item For the initialization stage, each $u\in V$ has its \emph{core numbers} initialized as its degree. Each $u$ will maintain a set of core numbers $u.core'$ of all neighbours $u.adj$ . Each $u$ sends its core numbers $u.core'$ to all neighbours.

    \item Each $u$ keeps receiving the core number from its neighbours $v\in u.adj$ and store the core numbers locally. Each node manages its own list of core numbers from neighbours.

    \item Once $u$ receives the core numbers from all neighbours, it executes Theorem~\ref{th:locality} to calculate its new core number. If the new core number is less than the old one, $u$ updates its core number with the new core number and sends it to its neighbours. Each $u$ does this step independently without waiting for other nodes to synchronize.

    \item Every time $u$ receives a new core number from its neighbours, it repeats Step 2 and Step 3 to recalculate its core number.

    \item The termination happens when all $u \in V$ finish calculating the core number and there is no new core number passing between any nodes.
\end{enumerate}

\begin{algorithm}[!htb]
\caption{Distributed Core Decomposition on Each Vertex $u\in V$ in Parallel}
\label{alg:go core-decompotion}
\SetAlgoNoEnd
\DontPrintSemicolon
\SetKwInOut{variable}{var}
\SetKwFunction{node}{node}
\SetKwFunction{updateCore}{updateCore}

%\variable{$allNodeChan: node \nrightarrow channel[node \times int]$}
\variable{\textit{node}: \textbf{string}}
\variable{\textit{allNodeChan} : \textit{node} $\nrightarrow$ \textbf{channel}[\textit{node} $\times$ \textbf{int}]}

%\SetKwProg{myproc}{process}{ \node{ \textit{u}: \textit{node}, \textit{neighbourChan}: \textit{node} $\nrightarrow$ \textbf{channel}[\textit{node} $\times$ \textbf{int}], \textit{selfChan}: \rm channel[\textit{node} $\times$ \textbf{int}]}}{}

\SetKwProg{myproc}{process}{ \textnormal{\textit{node}( \textit{u}: \textit{node}, \textit{neighbourChan}: \textit{node} $\nrightarrow$ \textbf{channel}[\textit{node} $\times$ \textbf{int}], \textit{selfChan}: \textbf{channel}[\textit{node} $\times$ \textbf{int}])}}{}
\myproc{}{

%\variable{$coreNumber: int \gets degree$}
\variable{\textit{coreNumber}: \textbf{int} $\gets$ \textbf{degree}}

\variable{\textit{storedNeighbourK}: \textit{node} $\nrightarrow$ \textbf{int} $\gets$ \{\}}
\variable{\textit{active}: \textbf{bool} $\gets$ \textit{true}}
\lFor{ \textit{v} $\in$ \textit{neighbourChan} }{  \textit{v}!$\langle$\textit{u, coreNumber}$\rangle$}

\While {\rm true} {
    \textit{selfChan}?$\langle$\textit{u, coreNumber}$\rangle$\;
    \textit{storedNeighbourK}[u] $\gets$ \textit{coreNumber}\;
    \If{$|$\textit{storedNeighbourK}$|$ $\geq$ \textit{degree}}{
        \textit{newCore} $\gets$ \updateCore(\textit{storedNeighbourK, coreNumber})\;
        \textit{active} $\gets$ \textit{true}\;
        \If{ \textit{newCore} $<$ \textit{coreNumber}}{
            \textit{coreNumber} $\gets$ \textit{newCore}\;
            \lFor {\textit{v} $\in$ \textit{adj}}{  \textit{v}!$\langle$u, \textit{coreNumber}$\rangle$}
            \textit{active} $\gets$ \textit{false}
            }
        }
    }
}
\SetKwProg{myproc}{function}{ \textnormal{\textbf{updateCore}(\textit{storedNeighbourK}: \textit{node} $\nrightarrow$  \textnormal{\textbf{int}}, \textit{coreNumber}: \textbf{int})} $\to$ \textbf{int}}{} 
\myproc{}{
\While{ $|$\textit{i} $\in$ \textit{storedNeighbourK} : \textit{i} $\geq$ \textit{coreNumber}$ |$ $<$ \textit{coreNumber}}{
\textit{coreNumber} $\gets$ \textit{coreNumber} – 1}
\Return \textit{coreNumber}
}
\end{algorithm}

\paragraph{Termination Detection.} 
Concurrent programming involves multiple processes that are executed simultaneously, often leading to complex interactions. Termination detection is a critical aspect of concurrent programming, ensuring that processes finish execution properly without deadlocks or infinite loops. There are several well-developed termination detection algorithms:
\begin{itemize}
    \item \emph{Chandy-Lamport Snapshot Algorithm}~\cite{chandy1985distributed}: It operates by taking snapshots of the local states of processes and the states of communication channels.
    \item \emph{Mattern’s Algorithm}~\cite{mattern1987algorithms}: It is an improvement over the Chandy-Lamport algorithm, which uses vector clocks to track causal relationships between events.
    \item \emph{Lai-Yang Algorithm}~\cite{lai1987distributed}: This algorithm is another approach that uses coloured markers (white and red) to capture global states without requiring $FIFO$ channels.
    \item  \emph{Dijkstra-Scholten Algorithm}~\cite{dijkstra1980termination}: This algorithm uses a hierarchical tree structure. The coordinator process at the root collects messages from child processes, which in turn collect messages from their children, and so on.
\end{itemize}

Instead of using a hierarchical tree structure like the {Dijkstra-Scholten Algorithm}~\cite{dijkstra1980termination}, we use a central server~\cite{montresordistributed} process and a dedicated channel called \emph{server channel} to collect messages from all nodes directly. This is a classic ~\emph{Master-Worker Paradigm}~\cite{van2002distributed} design. It involves a master node (central server) that coordinates the actions of the worker nodes (processes), which reports their status to the master. This concept is closely related to centralized monitoring and failure detection in distributed systems. It embodies the principles of centralized coordination~\cite{coulouris2005distributed}. Each node does not need to store or pass messages from other nodes. This approach is easier to implement and lightweight on the nodes.

Only \texttt{Active} nodes generate and send heartbeats to the server. \texttt{Inactive} nodes do not send heartbeat messages to the server because the server only needs to know if there are still \texttt{Active} nodes in the system. This approach greatly reduces the number of messages passed to the server.

The centralized approach is generally problematic for the distributed system because it introduces a single point of failure. This problem can be solved by adding a redundant server process as a backup. When the primary server process fails, the nodes will send $heartbeat$ to the backup server, and the backup server works the same as the primary.

\begin{itemize}
\item Initially, all nodes start with \texttt{Active} status since they will need to send their degree numbers to their neighbours. 
\item Once the \texttt{Active} nodes complete the calculation of $k$-core using the distributed $k$-core decomposition algorithm and send the $k$-core to their neighbours, they change the status to \texttt{Inactive}.
\item Inactive nodes can become active as soon as they receive updated core numbers from their neighbours because updated core numbers from their neighbours will trigger a recalculation of their own core number.
\end{itemize}

All nodes send heartbeat messages to the server under any of the following conditions:
\begin{enumerate}
    \item When the node receives an updated core number from its neighbour, it sets its status to active and immediately sends a heartbeat to the server.
    \item All \texttt{Active} nodes send heartbeats to the server every 10 seconds. If the $k$-core calculation takes too long, the node will use this periodic heartbeat to inform the \texttt{Active} status to the server. This prevents false termination from the server in the event that no heartbeat is received, but there are nodes that still calculate $k$-core numbers.
\end{enumerate}

The server Goroutine constantly reads messages from the server channel and checks if there are any incoming heartbeats. The server Goroutine does not store the status of each node; it uses a flag to indicate if it receives an \texttt{Active} heartbeat or not. This reduces the amount of memory used by the server. The server Goroutine periodically checks this flag. If there is no incoming heartbeat from any node for 5 minutes, the server will send a termination message to all nodes' termination channels to stop all nodes. The $k$-core number stored on each node is considered the final result. The 5-minute interval is much longer than the heartbeat interval of the node, which means that the server can capture the \texttt{Active} heartbeat from any node, even if the node is still calculating its core number. Since only the \texttt{Active} nodes send the heartbeat to the server and the server channel only adds a few milliseconds delay to the message passing, the 5-minute interval is sufficient to ensure the \texttt{Active} heartbeat will be received by the server and the server does not send any false termination signal. The heartbeat check interval can be longer than 5 minutes, but the downside of using a long check interval is that termination is delayed, and the system will not terminate as soon as it finishes calculating the core numbers. The detailed steps are shown in Algorithm~\ref{alg: termination detection}

\begin{algorithm}[!htb]
\caption{Termination Detection for Each Vertex $u\in V$ with WatchDog}
\label{alg: termination detection}
\SetAlgoNoEnd
\DontPrintSemicolon
\SetKwInOut{variable}{var}
\SetKwFunction{worker}{worker}
\SetKwFunction{watchdog}{watchdog}

\variable{\textit{heartbeat}: \textbf{channel[bool]}}
\variable{\textit{node}: \textbf{string}}
\variable{\textit{terminationChan}: \textit{node} $\nrightarrow$ \textbf{channel[bool]}}
\SetKwProg{myproc}{process}{ \textnormal{\textbf{worker}(\textit{u}: \textit{node}, \textit{selfTerminationChan}: \textbf{channel}[\textbf{bool}], \textit{heartbeat}: \textbf{channel}[\textbf{bool}])}}{}
\myproc{}{

\While{\textit{true}}{
    $\bold{if}\ … \to  \textit{heartbeat}!$\;
    $[]$ \textit{selfTerminationChan}? $\to \bold{return}$
}
}

\SetKwProg{myproc}{process}{ \textnormal{\textbf{watchdog}(\textit{heartbeat}: \textbf{channel}[\textbf{bool}], \textit{terminationChan}: \textit{node} $\nrightarrow$ \textbf{channel}[\textbf{bool}])}}{}
\myproc{}{

\textit{done} $\gets$ \textit{false}\;

\While{$\neg$ \textit{done}}{
    $\bold{if}\ \rm after(5 min)? \to$\;
       \quad\lFor {$\textit{c} \in \textit{terminationChan}$}{$\textit{c}!\textit{true}$}
       \quad$\textit{done} \gets \textit{true}$\;
    $[]$ \textit{heartbeat}? $\to$ \textit{skip}
}
}

\end{algorithm}

\paragraph{Algorithm with Termination Detection.}
We describe the combination of the distributed algorithm and the termination detection together in Algorithm~\ref{alg:go core-decompotion with termination detection}. 
Each node process uses a Go channel to receive messages from neighbours. This go channel must be asynchronous, and the channel capacity must be at least the number of neighbours. This is done to prevent node processes from sending and receiving messages at the same time. The function $GetCore$ performs the $k$-core decomposition calculation and returns the new core number to the node process. The node process will continue to run until it receives the termination signal from the termination channel. The termination signal is sent by the watchdog process. When a node process calculates the core number, it also sends a heartbeat message to the watchdog. If the watchdog process does not receive heartbeats within the past 5 minutes, it will broadcast a termination signal to all node processes.

\begin{algorithm}[!htb]
\caption{Combined Distributed Core Decomposition on Each Vertex $u\in V$ in Parallel with termination detection}
\label{alg:go core-decompotion with termination detection}
\SetAlgoNoEnd
\DontPrintSemicolon
\SetKwInOut{variable}{var}
\SetKwFunction{node}{node}
\SetKwFunction{updateCore}{updateCore}
\SetKwFunction{watchdog}{watchdog}

\variable{\textit{heartbeat}: \textbf{channel[bool]}}
\variable{\textit{node}: \textbf{string}}
\variable{\textit{allNodeChan}: \textit{node} $\nrightarrow$ \textbf{channel}[\textit{node} $\times$ \textbf{int}]}
\variable{\textit{terminationChan}: \textit{node} $\nrightarrow$ \textbf{channel[bool]}}

\SetKwProg{myproc}{process}{ \textnormal{\textit{node}(\textit{u}: \textit{node}, \textit{neighbourChan}: \textit{node} $\nrightarrow$ \textbf{channel}[\textit{node} $\times$ \textbf{int}], \textit{selfChan}: \textbf{channel}[\textit{node} $\times$ \textbf{int}], \textit{selfTerminationChan}: \textbf{channel}[\textbf{bool}], \textit{heartbeat}: \textbf{channel}[\textbf{bool}])}}{}
\myproc{}{

\variable{\textit{coreNumber}: \textbf{int} $\gets$ \textit{degree}}
\variable{\textit{storedNeighbourK}: \textit{node} $\nrightarrow$ \textbf{int} $\gets$ \{\}}
\variable{\textit{active}: \textbf{bool} $\gets$ \textit{true}}

\lFor{ $\textit{v} \in \textit{neighbourChan}$ }{ \textit{v}!$\langle$\textit{u, coreNumber}$\rangle$}

\While{\textit{true}}{
    \If {\textit{selfChan}\textnormal{?}$\langle$\textit{u, coreNumber}$\rangle$}{
        \textit{storedNeighbourK}[u] $\gets$ \textit{coreNumber}\;
        \If{$|$\textit{storedNeighbourK}$| \geq$ \textit{degree}}{
            \textit{new\_core} $\gets$ \updateCore(\textit{storedNeighbourK, coreNumber})\;
            \textit{active} $\gets$ \textit{true}\;
            \If{ \textit{new\_core} $<$ \textit{coreNumber}}{
                \textit{coreNumber} $\gets$ \textit{new\_core}\;
                \textit{heartbeat}!\;
                \lFor {\textit{v} $\in$ \textit{adj}}{ \textit{v}!$\langle$\textit{u, coreNumber}$\rangle$}
                \textit{active} $\gets$ \textit{false}
                }
            }
        }
    $[]$ \textit{selfTerminationChan}? $\to \bold{return}$
}
}

\SetKwProg{myproc}{process}{ \textnormal{\textbf{watchdog}(\textit{heartbeat}: \textbf{channel}[\textbf{bool}], \textit{terminationChan}: \textit{node} $\to$ \textbf{channel}[\textbf{bool}])}}{}
\myproc{}{

\textit{done} $\gets$ \textit{false}\;

\While{$\neg$ \textit{done}}{
    $\bold{if}\ \rm after(5 min)? \to$\;
       \quad\lFor {$\textit{c} \in \textit{terminationChan}$}{$\textit{c}!\textit{true}$}
       \quad$\textit{done} \gets \textit{true}$\;
    $[] \textit{heartbeat}? \to$ \textit{skip}
}
}

\SetKwProg{myproc}{function}{ \textnormal{\textbf{updateCore}(\textit{storedNeighbourK}: \textit{node} $\nrightarrow$  \textbf{int}, \textit{coreNumber}: \textbf{int})} $\to$ \textbf{int}}{}
\myproc{}{
\While{$|$\textit{i} $\in$ \textit{A} : \textit{i} $\geq$ \textit{coreNumber}$| <$  \textit{coreNumber}}{
$\textit{coreNumber} \gets \textit{coreNumber}$ – 1}
\Return \textit{coreNumber}
}
\end{algorithm}

\paragraph{Message Complexity.}
The performance of the distributed $k$-core decomposition algorithm can be measured using \emph{time complexity} or \emph{message complexity}~\cite{montresordistributed}. The time complexities are used to measure the total running time, while the message complexities are used to measure the number of messages passed between different nodes to complete the $k$-core decomposition. 
%%add citation from montresordistributed
The number of messages that pass through the go channels can be recorded during the program run time. It will not interfere with running the program locally or on a real distributed network. 
For distributed algorithms, since most of the running time is spent on the message passing through networks, which is much slower than accessing memory, we should mainly analyze the {message complexities}. 

We analyze the message complexities in the standard \emph{work-depth} model~\cite{jeje1992introduction}. 
The \emph{work}, denoted as $\mathcal W$, is the total number of operations that the algorithm uses.
The \emph{depth}, denoted as $\mathcal D$, is the longest chain of sequential operations. 
The work $\mathcal W$ is the total number of messages that the degree reduces to the core number, denoted as $\mathcal W = \mathcal O[\sum_{u\in V}{u.\Deg \cdot (u.\Deg - u.core)}]$, since the vertex must send messages to notify all neighbours when each time its core number decreases by one.

In the worst case, the process can be reduced to sequential running, e.g., a chain graph. In other words, the whole process needs the worst-case $n$ round to converge. Therefore, the depth $\mathcal D$ is equal to the work $\mathcal W$. 

\paragraph{Time Complexity.}
The time complexity of the algorithm is the same as the message complexity. Since it is a distributed environment, each node processes only the messages it receives. The sending and receiving of each message takes constant time $\mathcal O(1)$.
Furthermore, the computation after receiving one message is constant in time $\mathcal O(1)$. Therefore, the running time of the work and the depth are equal to the message complexity, that is $\mathcal W=\mathcal O[\sum_{u\in V}{u.\Deg \cdot (u.\Deg - u.core)}]$ and the depth $\mathcal D$ is equal to the work $\mathcal W$ in the worst case.

\subsection{Experimental Discussion}

In~\cite{mandal2017distributed}, Mandal et al.~design the distributed $k$-core decomposition algorithm running on the Spark cluster computing platform. 
In~\cite{aridhi2016distributed}, Aridhi et al.~propose distributed $k$-core decomposition and maintenance algorithms in large dynamic graphs, which are partitioned into multiple subgraphs and each of them assigned to a client.
In~\cite{montresordistributed}, a \emph{one-to-one} model was proposed in which each vertex in the graph represents one host. Simulations have been conducted with Peersim~\cite{peersim}. 
The experiment measures two metrics: the execution time and the total number of messages exchanged. The execution time is measured in rounds, that is, fixed-size time intervals during which each node has the opportunity to send one update message to all its neighbours. There are no mentions of the fixed-size time interval's duration, so we cannot determine the true running time. The average and maximum numbers of messages per node are generally comparable to the average and maximum degrees of nodes. Clearly, nodes with several thousand neighbours will be more overloaded than others.

Our experiments also focus on the \emph{one-to-one} model. Instead of using existing platforms such as Spark or Peersim, we use a native implementation in Golang. The experiments evaluate the total number of messages passed during the process of $k$-core decomposition. More specifically, we analyze the number of messages passed over different time intervals and how soon each node completes the $k$-core decomposition to obtain how the distributed $k$-core decomposition behaves over time. The measurement of the number of messages exchanged is comparable to the results in~\cite{montresordistributed}. The results of our experiment show similar behaviour: nodes with more neighbours will be more overloaded than others. 

 Termination detection is also discussed in ~\cite{montresordistributed}. There are three main approaches: 

\begin{itemize}
    \item $Centralized\ termination\ detection$, where each host notifies a central server when no new updates occur;
    \item $Decentralized\ termination\ detection$, which uses epidemic protocols for aggregation, allowing hosts to locally compute the global termination condition
    \item $Barrier\ synchronization$, where all nodes proceed synchronously, terminating when no further updates are exchanged.
    \item $Fixed\ number\ of\ rounds$, where the algorithm runs for a predetermined number of rounds, typically based on empirical results that suggest most real-world graphs converge in a small number of iterations.
\end{itemize}

Our experiment leverages a central server for termination detection, which is similar to the $centralized$  approach.

\section{Implementation} \label{implementation}

In this section, we discuss the implementation of the distributed $k$-core decomposition algorithm by simulating with Golang.
%The design logic of the program for algorithm simulation will be discussed.  

\subsection{Golang Simulation}

%\paragraph{Golang}
Golang, also known as Go, is a compiled programming language developed by Google \footnote{\url{https://go.dev/doc/}}. Go has built-in support for concurrent programming through \emph{Goroutines} and \emph{channels}. As lightweight threads, Goroutines are multiplexed onto a small number of operating system threads and are automatically scheduled by the Go runtime system. The Go channel allows Goroutines to send and receive data to and from each other safely and efficiently.

\paragraph{Why Choose Golang.}
In this paper, we try to simulate the distributed $k$-core decomposition algorithm. Since the number of exchanged messages remains constant with a distributed implementation, our simulation can be run on a multi-core machine instead of being executed in a truly distributed manner. Each vertex is a computational unit that is run by a lightweight thread. There can be millions of vertices in a tested data graph. Hence, our simulation has to run millions of lightweight threads in parallel. 
In addition, the vertices must exchange messages with one another. In other words, our simulation experiments require a programming language that efficiently supports a large number of concurrent lightweight threads that can be synchronized by message passing.

In addition to Go, there exist many other programming languages that support concurrent lightweight threads and message passing, such as Erlang\footnote{\url{https://www.erlang.org/docs}}, 
Haskell\footnote{\url{https://www.haskell.org/}},
Elixir\footnote{\url{https://elixir-lang.org/}}, and Rust\footnote{\url{https://www.rust-lang.org/}}.
%%% ADD perf benchmark between these
A study~\cite{6920368} shows that concurrency in Go is easier to implement and has better performance than Java.

\subsection{Algorithm Implementation}

The simulation program \emph{distributed-k-core.go}\footnote{\url{https://github.com/Marcus1211/MEng/blob/main/implementation/distributed-k-core.go}} contains the following functions:
\begin{itemize}
    \item \emph{receive}: Each node continuously runs the receive functions until the end. This function receives the incoming message from the neighbours and calculates the node's $k$-core.
    \item \emph{send}: Once the \emph{receive} function calls this function to send $k$-core to neighbours after calculating the node's $k$ core number.
    \item \emph{updateCore}: The \emph{receive} function calls this function to calculate the node's $k$-core number after receiving the $k$-core number from its neighbours.
    \item \emph{sendHeartBeat}: Each node uses this function to send $heartbeat$ to the central server. The message sent to the server is called a \emph{{heartbeat}}. It is a message that each node sends to the server to inform its status.
    \item \emph{receiveHeartBeat}: The central server continuously runs this function to receive $hearbeat$ messages from all nodes and determine whether the termination signal should be issued.
    \item \emph{dataCleanse}: This function processes the graph data to make it usable for the simulation.
\end{itemize}

\paragraph{Data Structure.} \label{node data structure}
%%% How program runs, terminations, message passing, latency, etc...
The following is the implementation of the node. Each node maintains the following variables by itself and updates values by exchanging information with its neighbours; 

\begin{itemize}
    \item ID: Each node has its own unique ID as an identifier.
    \item storedNeighbourK: Each node stores the core number of its neighbours.
    \item active: This boolean value indicates the status of the node. The value $True$ refers to the status of $active$, and the value $False$ refers to the status of $inactive$. 
    \begin{itemize}
    \item \texttt{Active}: The node is actively calculating its $k$-core number and passing messages to its neighbours
    \item \texttt{Inactive}: The node is currently not processing passed messages or calculating its $k$-core number.
    \end{itemize}
    \item selfChan: Each node has its own channel to receive messages from neighbour nodes.
    \item heartbeat: The channel used by the server to receive heartbeat from all nodes.
    \item selfTerminationChan: Each node has a channel to receive the termination messages from the server.
    \item neighbourChan: Each node stores its neighbour channels.
\end{itemize}

The message passed between nodes has the following structure, where $ID$ is the same as in the node data structure and $coreNumber$ is the current core number of the node.
\begin{lstlisting}[language=Go]
        type sendMsg struct {
            ID   string
            coreNumber int
        }
\end{lstlisting}

\begin{figure}[!htb]
    \centering
    \includegraphics[width=1\linewidth]{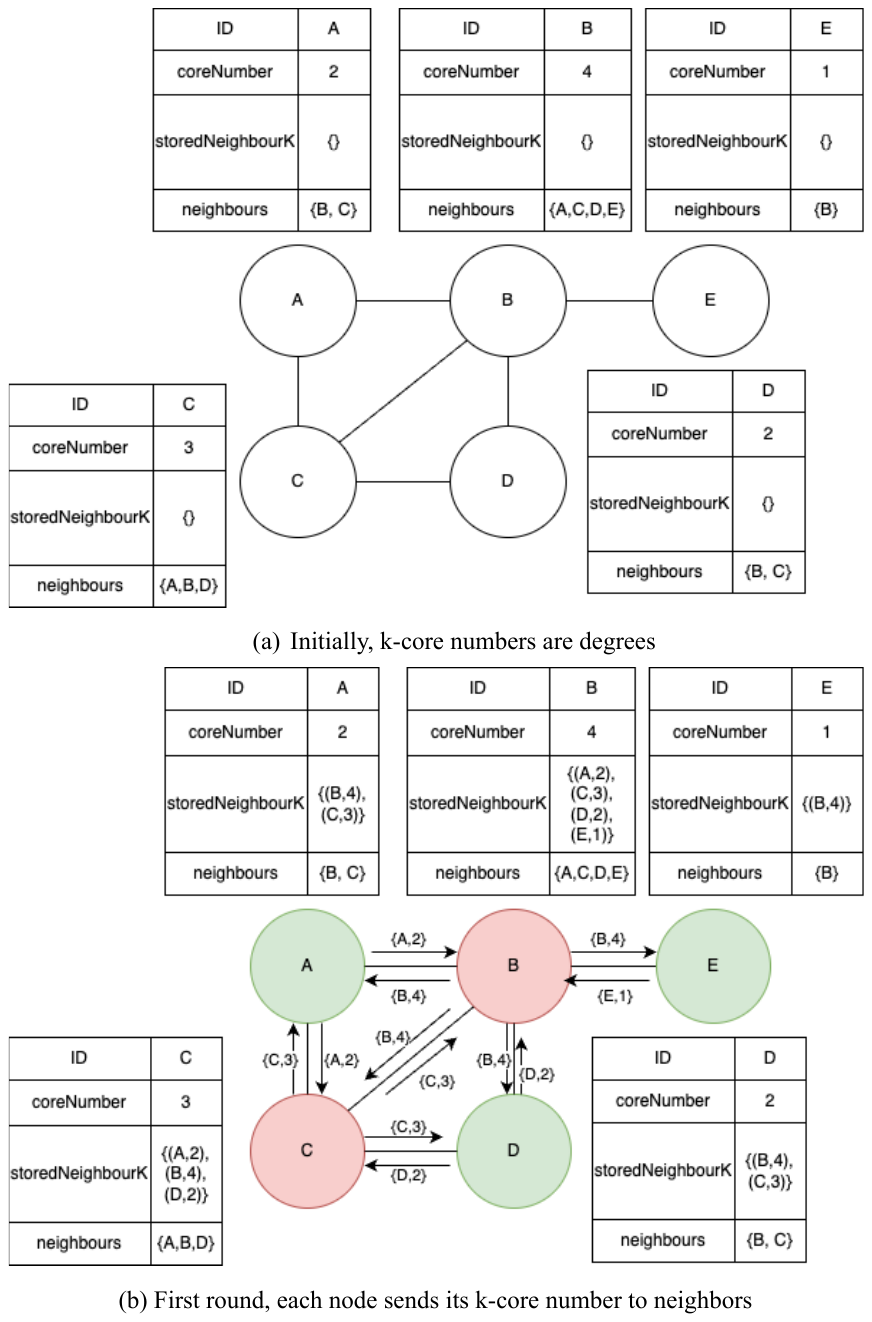}
    \caption{Example of distributed $k$-core decomposition with message passing (a)}
    \label{fig:decom sample (a)}
\end{figure}

\begin{figure}[!htb]
    \centering
    \includegraphics[width=0.8\linewidth]{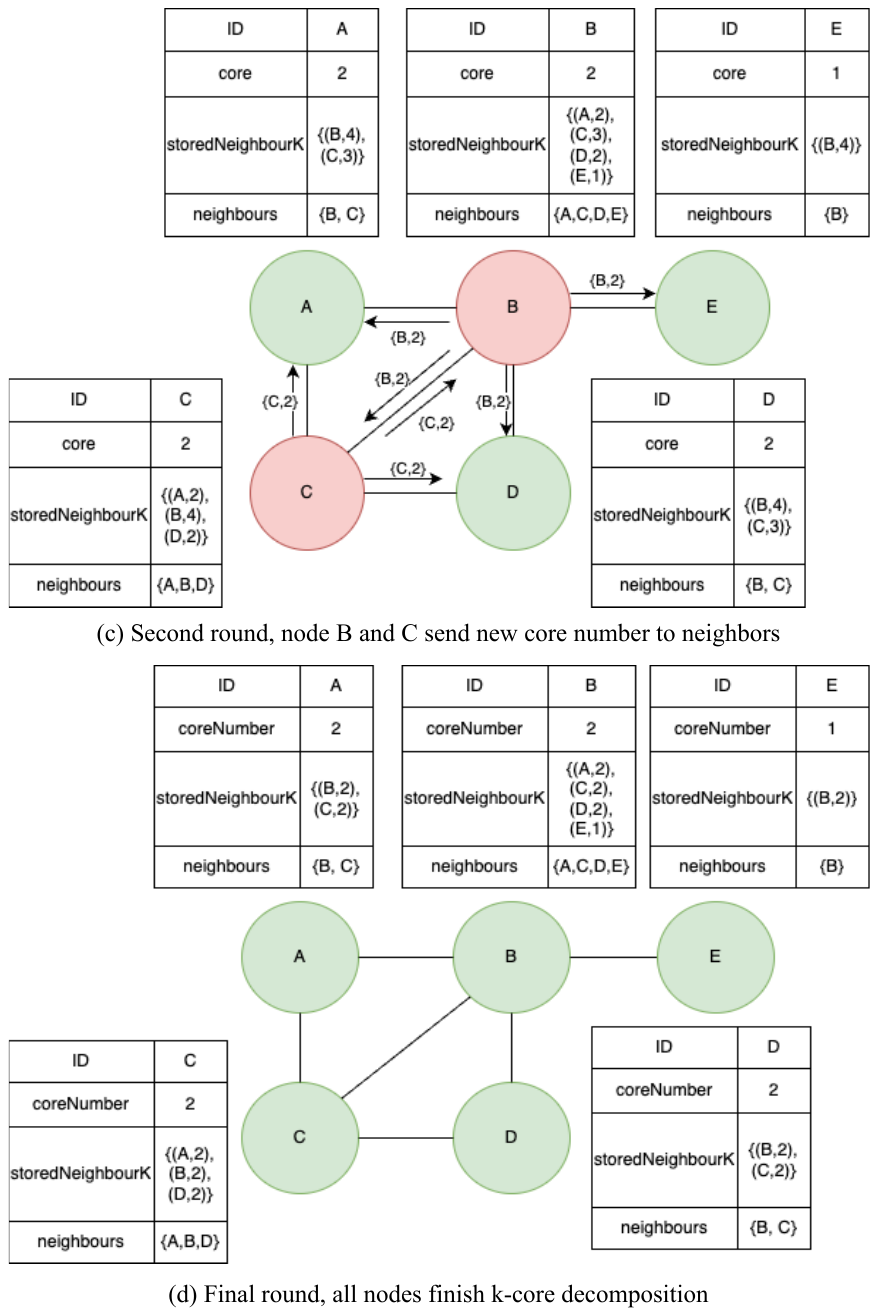}
    \caption{Example of distributed $k$-core decomposition with message passing (b)}
    \label{fig:decom sample (b)}
\end{figure}

\paragraph{Message Passing.}
Figures \ref{fig:decom sample (a)} and \ref{fig:decom sample (b)} show the distributed $k$ core decomposition procedure with a small example, where the \texttt{Active} nodes are coloured red, the \texttt{Inactive} nodes are colored green, and the arrows between the nodes demonstrate how messages are passed. 
The side table indicates the node data structure in Section~\ref{node data structure} stored in each node. 
%, where $ID$ is the identifier of the node, $coreNumber$ is the core number of the node, $storedNeighbourK$ is the neighbours' $coreNumber$, and $neighbours$ are the connected nodes. 
The message contains the sender's ID and core number, for example, the message $\{A, 2\}$, in which $A$ is the ID and $2$ is the core number.

\begin{itemize}
    \item Initially, as shown in Figure~\ref{fig:decom sample (a)}(a), all nodes are initialized with degree number as their $coreNumber$; all nodes are aware of their neighbours but do not know the $coreNumber$ of their neighbours
    
    \item As shown in Figure~\ref{fig:decom sample (a)}(b), all nodes send their $coreNumber$ to all neighbours. Nodes $A$, $D$, and $E$ do not need to decrease their $coreNumber$. Nodes $B$ and $C$ decrease their $coreNumber$ to $2$ according to the distributed $k$-core decomposition algorithm.
    
    \item As shown in Figure~\ref{fig:decom sample (b)}(c), nodes $B$ and $C$ send a new $coreNumber$ to their neighbours.
    \item Finally, as shown in Figure~\ref{fig:decom sample (b)}(d), all nodes have updated $coreNumber$ from nodes $B$ and $C$; no nodes need to update their $coreNumber$; all nodes enter the \texttt{Inactive} status.
\end{itemize}

\paragraph{Testing.}
In this work, the black-box testing is used to ensure the accuracy of the distributed $k$-core decomposition algorithm. Black-box testing is a software testing method in which the internal structure, design, and implementation of the program being tested are not known. The test focuses solely on the inputs and outputs of the program. Since the BZ Algorithm~\cite{bz2003} is a well-recognized algorithm for $k$-core decomposition, it is used as the reference for black-box testing. There is another program \emph{bz-origin.go}\footnote{\url{https://github.com/Marcus1211/MEng/blob/main/test/bz-origin.go}} to implement the sequential BZ Algorithm~\cite{bz2003} to verify the final results. We run the BZ Algorithm against all test graph data in this experiment to obtain the core number of each vertex; then we compare the core numbers of each vertex from both the BZ Algorithm and the distributed $k$-core decomposition algorithm to make sure both algorithms produce the same $k$-core decomposition results.

The graph data used in this experiment are from the real world and are often quite large. It is difficult to use large graph data for testing purposes, as it takes more time to obtain the testing results. In addition, it is difficult to guarantee the correctness of both the BZ algorithm program and the distributed $k$-core decomposition program. Hence, we used some synthetic graph data to test some edge cases of the graphs and the correctness of the programs. Since the synthetic test graph data are small, we manually perform the $k$-core decomposition to obtain the core number for each vertex. The result of manual core decomposition is verified and used as a reference for both the BZ Algorithm and the distributed $k$-core decomposition algorithm programs.
Test cases of the synthetic graph data include:
\begin{itemize}
    \item Circle graph: All nodes in this graph form a circle.
    \item Linear graph: All nodes in this graph form a linear line.
    \item Complete graph: Undirected graph in which every pair of distinct vertices is connected by a unique edge.
    \item Small graph: This is the normal graph that can be used for regular testing purposes.
    \item Incomplete graph: 
        \begin{itemize}
            \item Missing edge: Some of the edges are missing from this graph. e.g., Node A has the edge connection to B but Node B does not have the connection to A. This can be used to test directed graphs.
            \item Missing node: Some of the nodes are missing in this graph. e.g., Node A has the connection to Node B, but Node B is missing in these graph data.
        \end{itemize}
\end{itemize}

\section{Experiments} \label{experiment}

In this section, we carry out extensive experiments to evaluate the message complexity of the distributed $k$-core decomposition algorithm with the simulation results for each data graph. The analysis contains the number of total passing messages, the number of passing messages over time interval, and the number of \texttt{Active} nodes over time interval.

\subsection{Experiment Setup}

The experiments are performed on a server with an AMD CPU (64 cores, 128 hyperthreads, 256 MB of last-level shared cache) and 256 GB of main memory. The server runs the Ubuntu Linux (22.04) operating system. The distributed $k$-core decomposition algorithm is implemented in Golang (1.21).
By default, Golang leverages all cores of the machine. However, based on different setups for the experiment, Golang may only use a single core for the experiment. The experiment code prints out the number of computing cores used to execute the program. If a single core is used because of the old version of GO, researchers need to add the following snippet to the code $main$ function\footnote{\url{https://github.com/Marcus1211/MEng/blob/main/implementation/distributed-k-core.go\#L156}}
\begin{lstlisting}[language=Go]
    runtime.GOMAXPROCS(runtime.NumCPU())
\end{lstlisting}
The source code implementation is shared on the Github repository\footnote{\url{https://github.com/Marcus1211/MEng}}.

\subsection{Tested Graphs}
We select 14 real-world graphs within seven categories.
All graphs are obtained from the Stanford Large Network Dataset Collection (SNAP)\footnote{\url{https://snap.stanford.edu/data/}}, shown in Table~\ref{Tested graphs}.  
%As stated in the~\cite{snap}, SNAP dataset is a general-purpose, high-performance system that provides easy-to-use high-level operations for the analysis and manipulation of large networks. 
SNAP~\cite{snap} is a widely used repository of large real-world networks, providing easy-to-use formats and descriptions.
It is widely used in graph-related research, including~\cite{montresordistributed},~\cite{snap2}, and~\cite{snap3}. The detailed graph descriptions are summarized as follows.
\begin{itemize}
    \item soc-pokec-relationships(SPR): The most popular social network on the Internet in Slovakia.
    \item musae-PTBR-features(PTBR): Twitch user-user networks of gamers who stream in a certain language.
    \item facebook-combined(FC): This data set consists of \texttt{"}circles\texttt{"} or \texttt{"}friends lists\texttt{"} from Facebook.
    \item musae-git-features(MGF): A large social network of GitHub developers that was collected from the public API in June 2019.
    \item soc-LiveJournal1(LJ1): A free on-line community with almost 10 million members; a significant fraction of these members are highly active.
    \item email-Enron(EEN): The Enron email communication network covers all email communication within a dataset of around half a million emails.
    \item email-EuAll(EEU): The network was generated using email data from a large European research institution.
    \item p2p-Gnutella31(G31): A sequence of snapshots of the Gnutella peer-to-peer file-sharing network from August 2002.
    \item com-lj(CLJ): LiveJournal friendship social network and ground-truth communities.
    \item com-amazon(CA): Ground-truth community defined by product category provided by Amazon.
    \item web-Stanford(WS): The nodes represent pages from Stanford University (stanford.edu), and the directed edges represent the hyperlinks between them.
    \item web-Google(WG): Nodes represent web pages, and directed edges represent hyperlinks between them
    \item amazon0505(A0505): The network was collected by crawling the Amazon website.
    \item soc-Slashdot0811(S0811): The network contains friend-foe links between Slashdot users.
\end{itemize}

%%%%pre process graph

Each graph is pre-processed before the experiment. For simplicity, all directed graphs are converted to undirected graphs based on the following rules:
\begin{itemize}
    \item  A vertex can not connect to itself.
    \item Each pair of vertices can only connect with one edge.
    \item All graphs are converted into JSON format, in which the key is the vertex, and all its neighbour vertices are stored as list of numbers in the value.
\end{itemize}

\paragraph{Graph Properties.}
In Table~\ref{Tested graphs}, we can see that the tested graphs have up to millions of edges. Their average degrees range from 2 to 46, and their maximal core numbers range from 8 to 376. Each column in Table 1.1 is explained below.
\begin{itemize}
    \item Type: Tested graphs have two types, directed and undirected. Directed graphs contain directed edges, which means that connections are not mutual between connected nodes. Undirected graphs have edges that do not have a direction. Directed graphs must be converted into undirected graphs before the experiment.
    \item $n$ = $|V|$: the number of vertices in the graph; $n$ is the number of nodes in the experiment, which is the same as $V$.
    \item $m$ = $|E|$: $E$ is the number of edges in the graph.
    \item Avg-Degree: The number of neighbours of a node is called the degree; the average degree is the sum of all degrees divided by the total number of nodes.
    \item Max-Degree: The maximum degree among all nodes.
    \item Max core: The maximum number of cores among all nodes after $k$-core decomposition.
\end{itemize}

%%%use short-name in figure discussion and \emph

\begin{landscape}
\begin{table}[!htb]
\centering
\begin{tabular}{|p{30mm}|l|rrrrrr|}
        \hline
          Category&Graph  &Type&  $n = |V|$&  $m = |E|$& Avg-Degree&Max-Degree&Max-Core\\
         \hline
          Social Networks&soc-pokec-relationships(SPR)&Directed&  1,632,803&  30,622,564	& 29  &14739&118\\
          &musae-PTBR-features(PTBR) &Undirected&  1,912&  31,299& 24  &1635&21\\
          &facebook-combined(FC) &Undirected&  4039&  88234& 46  &986&118\\
          &musae-git-features(MGF) &Undirected&  37,700&  289,003& 36  &28191&29\\
          &soc-LiveJournal1(LJ1) &Directed&  4,847,571&  68,993,773& 19  &20314&376\\
         \hline
          Communication &email-Enron(EEN) &Undirected&  36,692&  183,831& 10  &1383&49\\
          \hline
          Networks &email-EuAll(EEU) &Directed&  265,214&  420,045& 2  &7631&44\\
         \hline
          Internet \newline peer-to-peer \newline networks&p2p-Gnutella31(G31) &Directed&  62,586&  147,892& 7  &68&9\\
        \hline
          Networks with \newline Ground-Truth Communities&com-lj(CLJ) &Undirected& 3,997,962& 34,681,189&25  &14208&360\\
          &com-amazon(CA) &Undirected& 334,863& 925,872&5  &546&8\\
        \hline
          Web Graphs&web-Stanford(WS) &Directed& 281,903& 2,312,497&14  &38625&75\\
          &web-Google(WG) &Directed& 875,713& 5,105,039&10  &6331&44\\
        \hline
          Product \newline Co-Purchasing \newline Networks&amazon0505(A0505) &Directed& 410,236& 3,356,824&12  &2760&15\\
        \hline
          Signed Networks&soc-Slashdot0811(S0811) &Directed& 77,357& 516,575&13  &2540&59\\
        \hline
    \end{tabular}
    \caption{Tested Data Graphs}
    \label{Tested graphs}
\end{table}
\end{landscape}

\begin{figure}[!htb]
    \centering
    \includegraphics[width=0.8\linewidth]{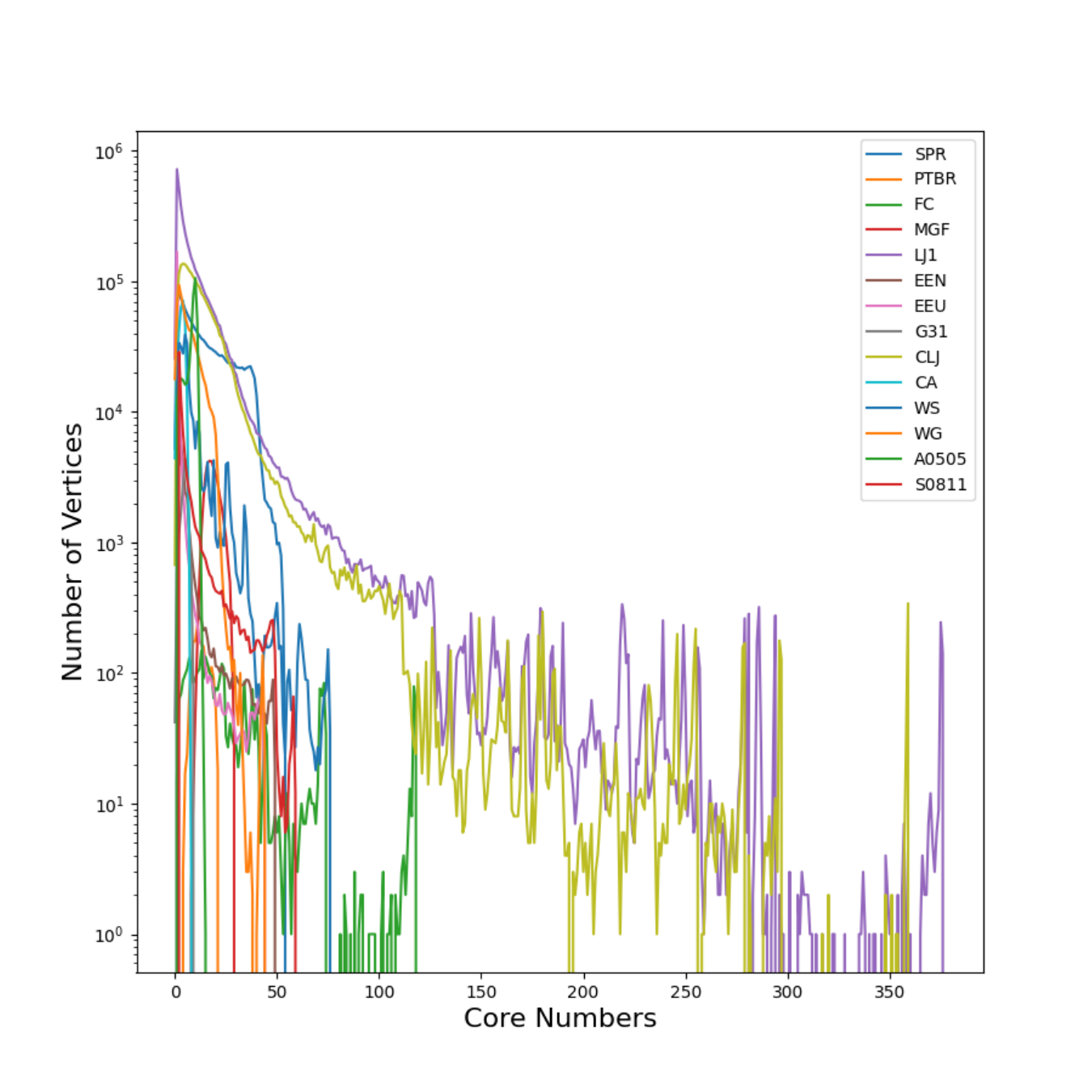}
    \caption{Core Number Distribution}
    \label{fig:Core Number Distribution}
\end{figure}

\paragraph{Core Number Distribution.}
In Figure~\ref{fig:Core Number Distribution}, we can see that the core numbers of vertices are not uniformly distributed in all the graphs tested, where the x-axis is core numbers and the y-axis is the number of vertices. That is, a great portion of vertices have small core numbers, and few have large core numbers. This indicates that the worst case in the graph data does not appear in practice because all test graphs are from the real world, and none of them have a large distribution of high core numbers. For example, \emph{LJ1} has 0.5 million vertices with a core number of ~$1$; \emph{PTBR} and \emph{MGF} have no vertices with a core number of $1$. Although \emph{CA} has more than $300000$ nodes, all core numbers of vertices range from $0$ to $8$. 

%\subsection{Message Complexity Evaluation}

\subsection{Evaluate Total Number of Messages}

\begin{figure}[!htb]
    \centering
    \includegraphics[width=0.8\linewidth]{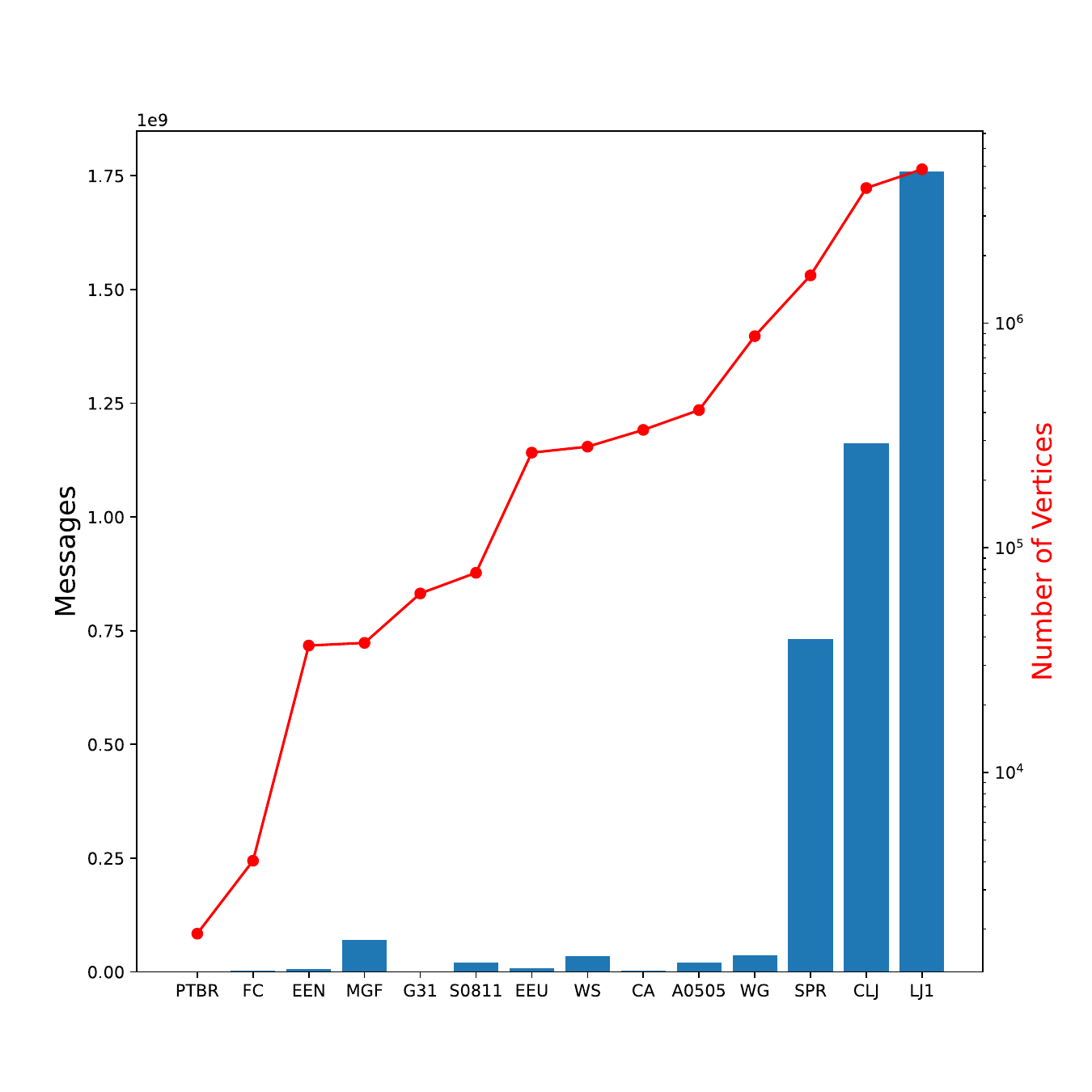}
    \caption{Total Number of Passing Messages}
    \label{fig:Total Number of Passing Messages}
\end{figure}

Figure~\ref{fig:Total Number of Passing Messages} shows the total number of passing messages for each graph, where the x-axis is the tested graphs and the y-axis is the total number of passing messages for each graph to complete the distributed $k$-core decomposition. Generally, large graphs, such as \emph{SPR} and \emph{LJ1}, with more nodes and edges, require passing more messages to complete the simulation.

However, the average degree can also affect the number of messages passed, even when the number of nodes is small. For example, \emph{MGF} has $37,700$ nodes with an average degree of $36$, and \emph{FC} has only $4039$ nodes but with an average degree of $46$. Both \emph{MGF} and \emph{FC} are considered small graphs but require a large number of messages to complete the simulation. The high average degree means that each node has more neighbours on average. When a node updates its core number, it needs to send more messages due to more neighbours. Hence, more messages are passed. The number of messages passed does not increase linearly with the average degree. It also depends on the core distribution among all nodes.

\subsection{Evaluate the Number of Message Passing with Time Interval}

\begin{figure}[!htb]
    \centering
    \includegraphics[width=0.8\linewidth]{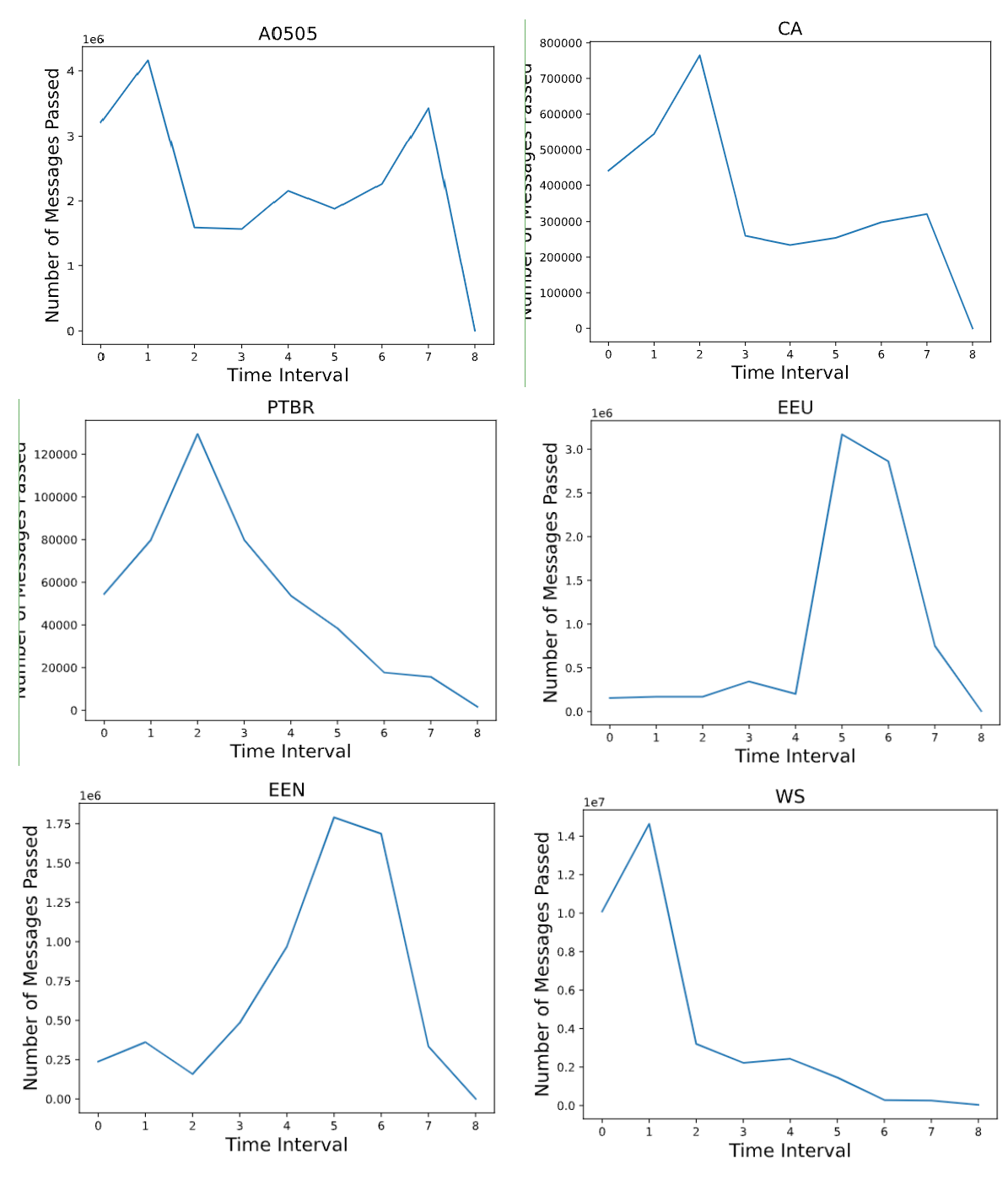}
    \caption{Number of Messaged Passed over Time Interval (Part 1)}
    \label{fig:Number of Messaged Passed over Time Interval (a)}
\end{figure}

\begin{figure}[!htb]
    \centering
    \includegraphics[width=0.8\linewidth]{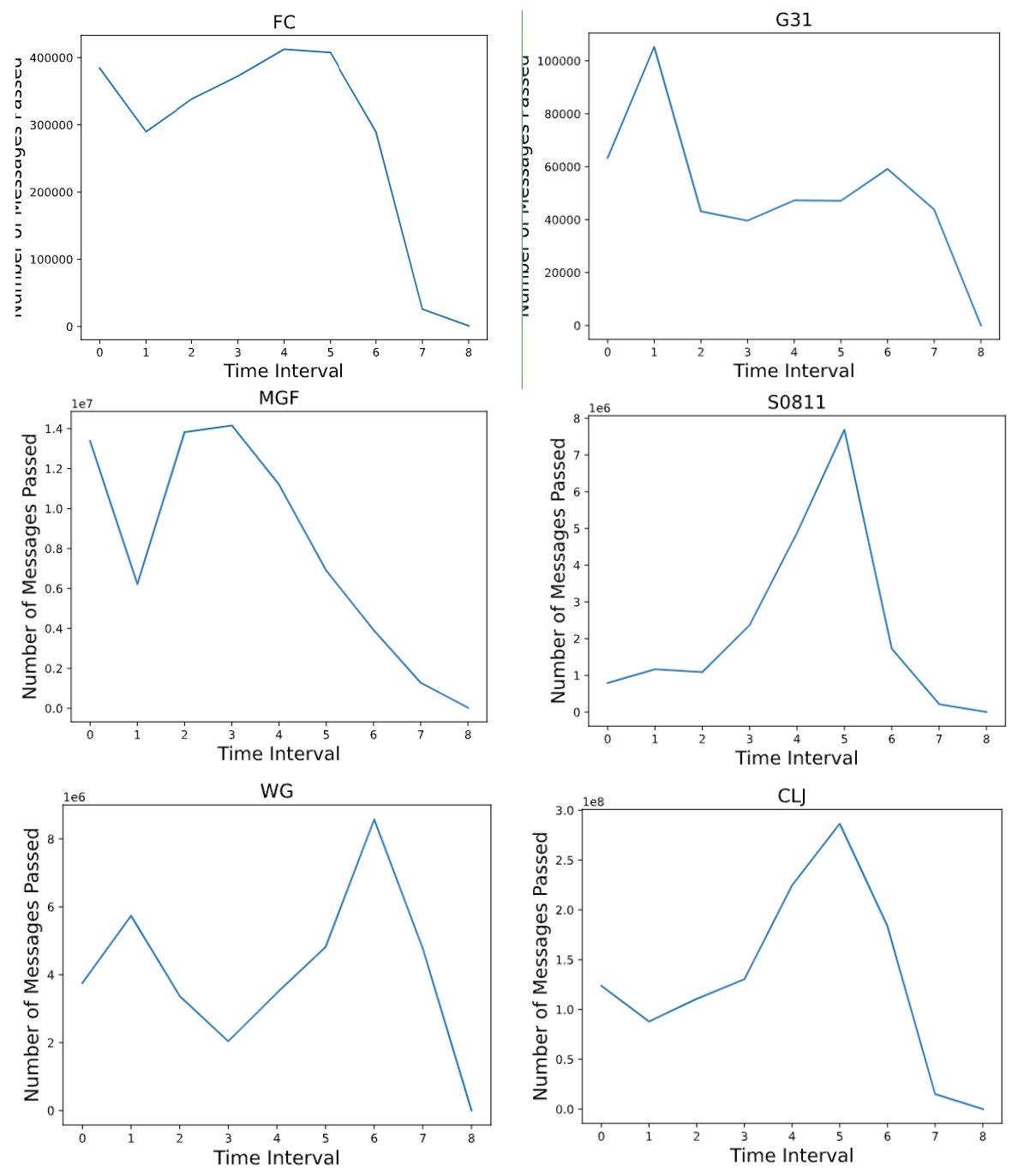}
    \caption{Number of Messaged Passed over Time Interval (Part 2)}
    \label{fig:Number of Messaged Passed over Time Interval (b)}
\end{figure}

\begin{figure}[!htb]
    \centering
    \includegraphics[width=0.8\linewidth]{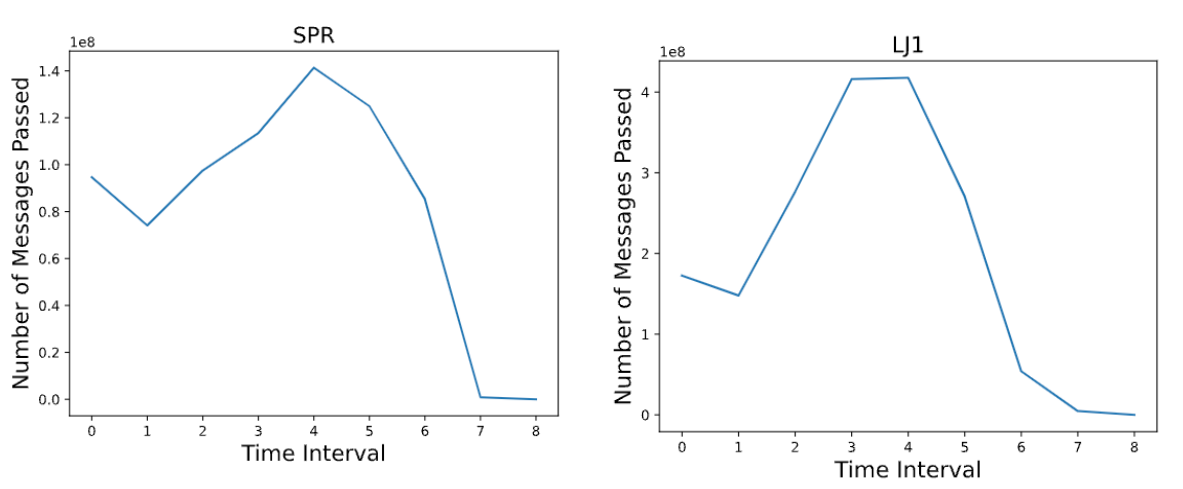}
    \caption{Number of Messaged Passed over Time Interval (Part 3)}
    \label{fig:Number of Messaged Passed over Time Interval (c)}
\end{figure}

Figures~\ref{fig:Number of Messaged Passed over Time Interval (a)}, ~\ref{fig:Number of Messaged Passed over Time Interval (b)} and ~\ref{fig:Number of Messaged Passed over Time Interval (c)} show the number of messages passed in the system over time, where the Y axis is the number of total messages passed between all nodes when simulating the distributed core-decomposition algorithm and the X axis is an abstract time interval. We captured the overall running time of each experiment and equally divided it into eight time slices; then we collected the number of messages passed in each time slice to obtain the data points to plot the graph. Since different graphs took different times to complete the experiment; Some large graphs take up to 15 hours to complete, using abstracted time intervals instead of real time to plot the graph helps demonstrate the general trend of each simulation. 

As shown in Figures \ref{fig:Number of Messaged Passed over Time Interval (a)},~\ref{fig:Number of Messaged Passed over Time Interval (b)} and ~\ref{fig:Number of Messaged Passed over Time Interval (c)}, most of the messages are passed in the first couple of time intervals. This is expected as all nodes need to share their degree number with their neighbour when initiating the core decomposition. Every time a node updates its core number, it needs to pass messages to its neighbours; Most of the nodes remain in the active state and keep updating their core number at the beginning of the simulation. Hence, a large amount of messages is passed during the first couple of time intervals. The number of messages passed decreases for all graphs as the simulation progresses because more nodes finish calculating their core number. Hence, no more messages are sent from the inactive node.

The graph \emph{WG} demonstrates a special case in which there is a spike in messages in the middle of the simulation. This is due to the core calculation of nodes that have many neighbours; When they update their core number, they send messages to a large number of neighbours, which causes neighbours to update their core numbers as well. Hence, a recursive effect is caused, which causes a spike in the number of messages passed. This shows the correlation with the distribution of the core number in Figure \ref{fig:Core Number Distribution}. The graph $WG$ has more than 800000 nodes, but the maximum core number is 44, which means that a large number of nodes have the same core number. These nodes need to calculate their core number and send the calculated core number to their neighbours. This causes a spike in the number of messages passed.

\subsection{Evaluate Number of Active Nodes for Each Time Interval}

\begin{figure}[!htb]
    \centering
    \includegraphics[width=0.8\linewidth]{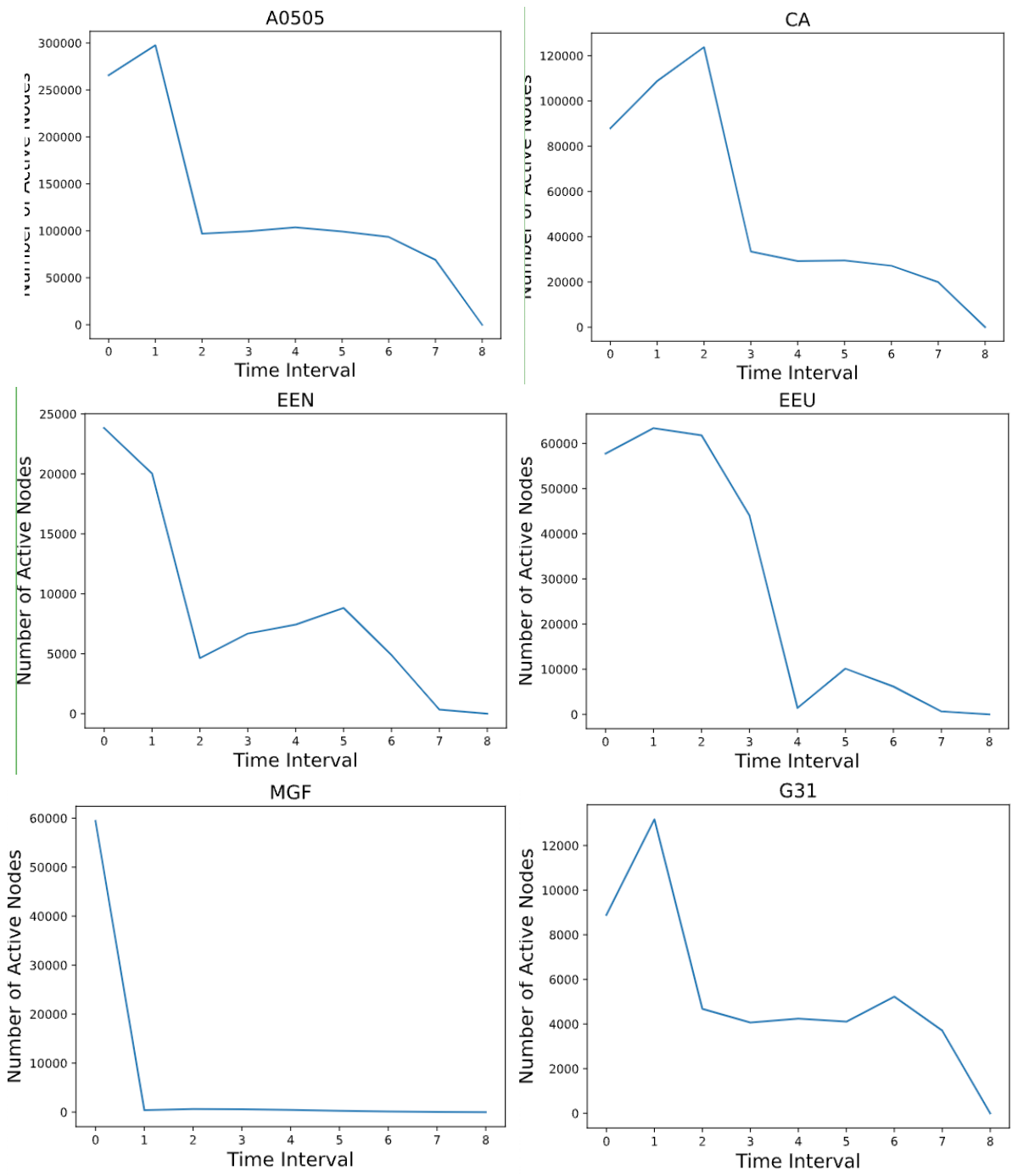}
    \caption{Number of Active Node over Time Interval (Part 1)}
    \label{fig:Number of Active Node over Time Interval (a)}
\end{figure}

\begin{figure}[!htb]
    \centering    
    \includegraphics[width=0.8\linewidth]{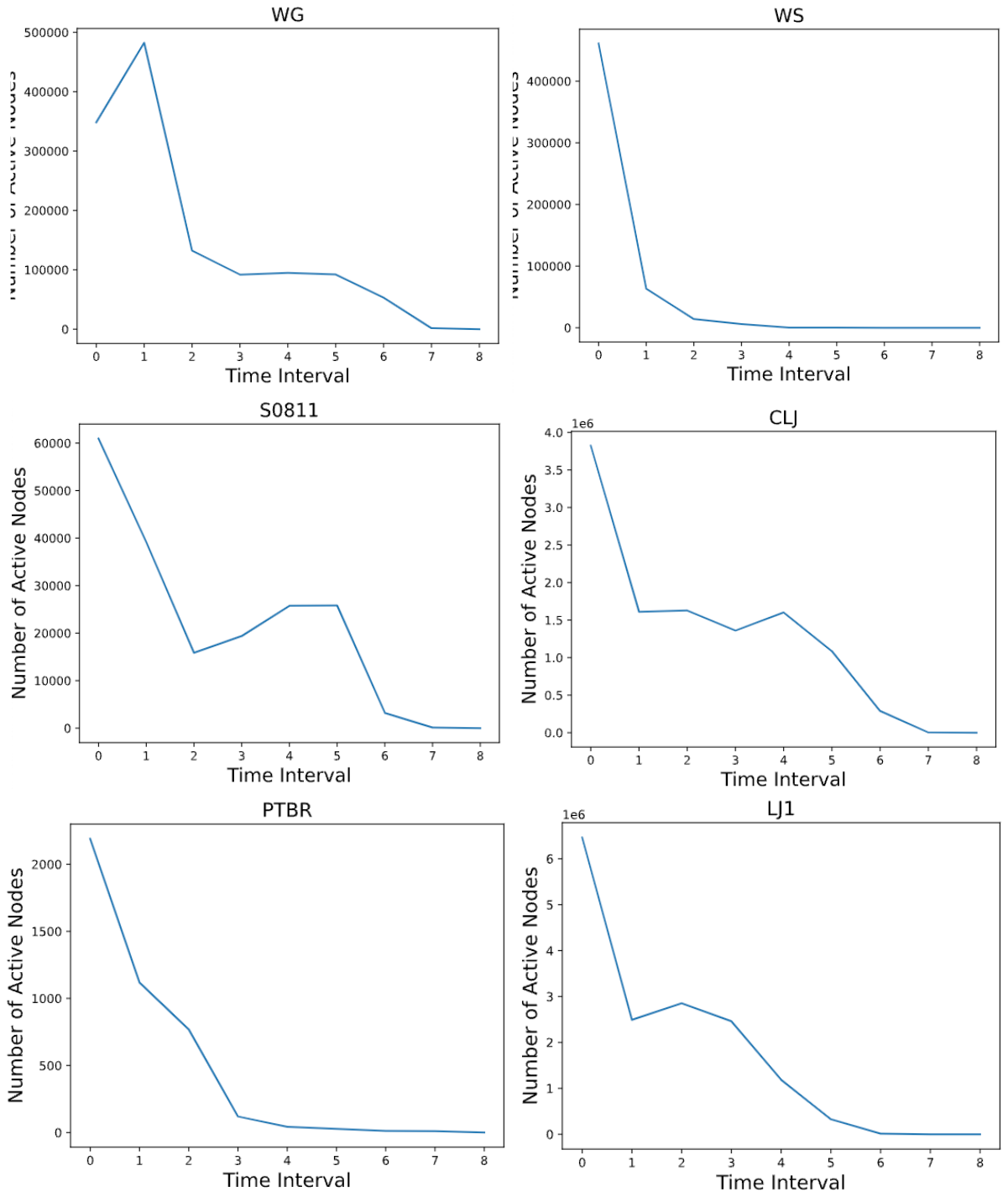}
    \caption{Number of Active Node over Time Interval (Part 2)}
    \label{fig:Number of Active Node over Time Interval (b)}
\end{figure}

\begin{figure}[!htb]
    \centering    
    \includegraphics[width=0.8\linewidth]{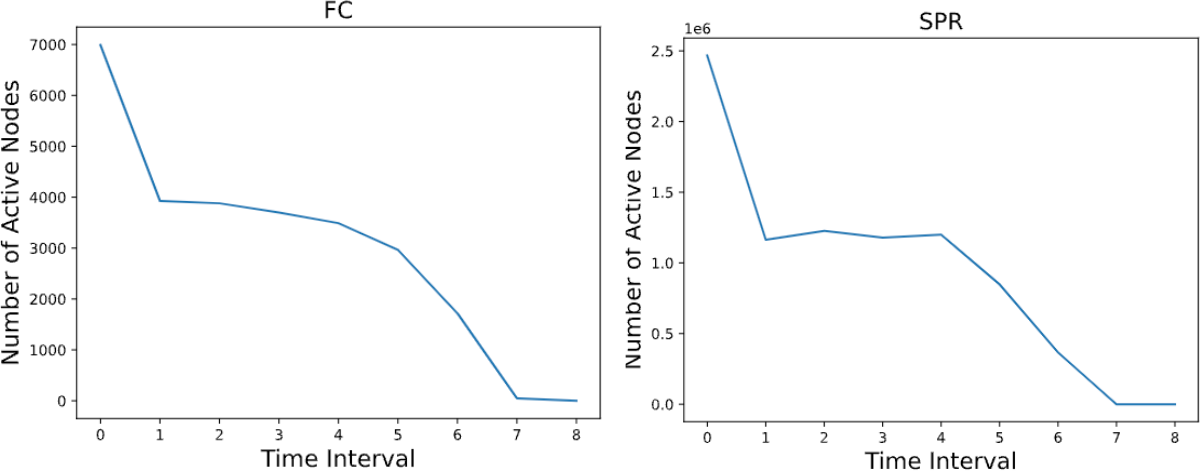}
    \caption{Number of Active Node over Time Interval (Part 3)}
    \label{fig:Number of Active Node over Time Interval (c)}
\end{figure}
Figures~\ref{fig:Number of Active Node over Time Interval (a)}, ~\ref{fig:Number of Active Node over Time Interval (b)} and ~\ref{fig:Number of Active Node over Time Interval (c)} show the number of \texttt{Active} nodes over time where the x-axis is the same abstract time interval shown in Figures~\ref{fig:Number of Messaged Passed over Time Interval (a)}, ~\ref{fig:Number of Messaged Passed over Time Interval (b)} and ~\ref{fig:Number of Active Node over Time Interval (c)} and ~y-axis is the number of total messages passed between all nodes. Most of the nodes are in the \texttt{Active} state at the beginning of the simulation. As stated in the implementation, the node will only enter the \texttt{Active} state when it needs to decrease its core number and send messages to its neighbour. Hence, more and more nodes enter the inactive state once they finish calculating their core number and do not receive messages from their neighbours to trigger a further core number calculation. The number of \texttt{Active} nodes decreases at different rates despite running on the same machine. This is caused by the distribution of nodes with different core numbers. For some graphs such as \emph{A0505} or \emph{EEN}, most of the nodes have small core numbers, so they will be processed quickly and enter the \texttt{Inactive} state. On the other hand, graphs like \emph{CLJ} or \emph{SPR}, have more nodes with a higher core number; It will take some time for them to process these nodes; therefore, the number of \texttt{Active} nodes does not drop rapidly at the beginning of the simulation.

The speed of the distributed $k$-core decomposition algorithm is determined by the number of \texttt{Active} nodes remaining in the experiment. As the experiment progresses, more and more nodes turn into the \texttt{Inactive} state, and eventually, all nodes become \texttt{Inactive}. Nodes with a small core number always turn inactive first, which means if the speed of the algorithm is determined by the core number distribution among all nodes, e.g., if most of the nodes have a high core number, it would take a longer time for the distributed $k$-core decomposition to finish.

\subsection{Evaluate the Total Running Time}

\begin{figure}[!htb]
    \centering
    \includegraphics[width=0.8\linewidth]{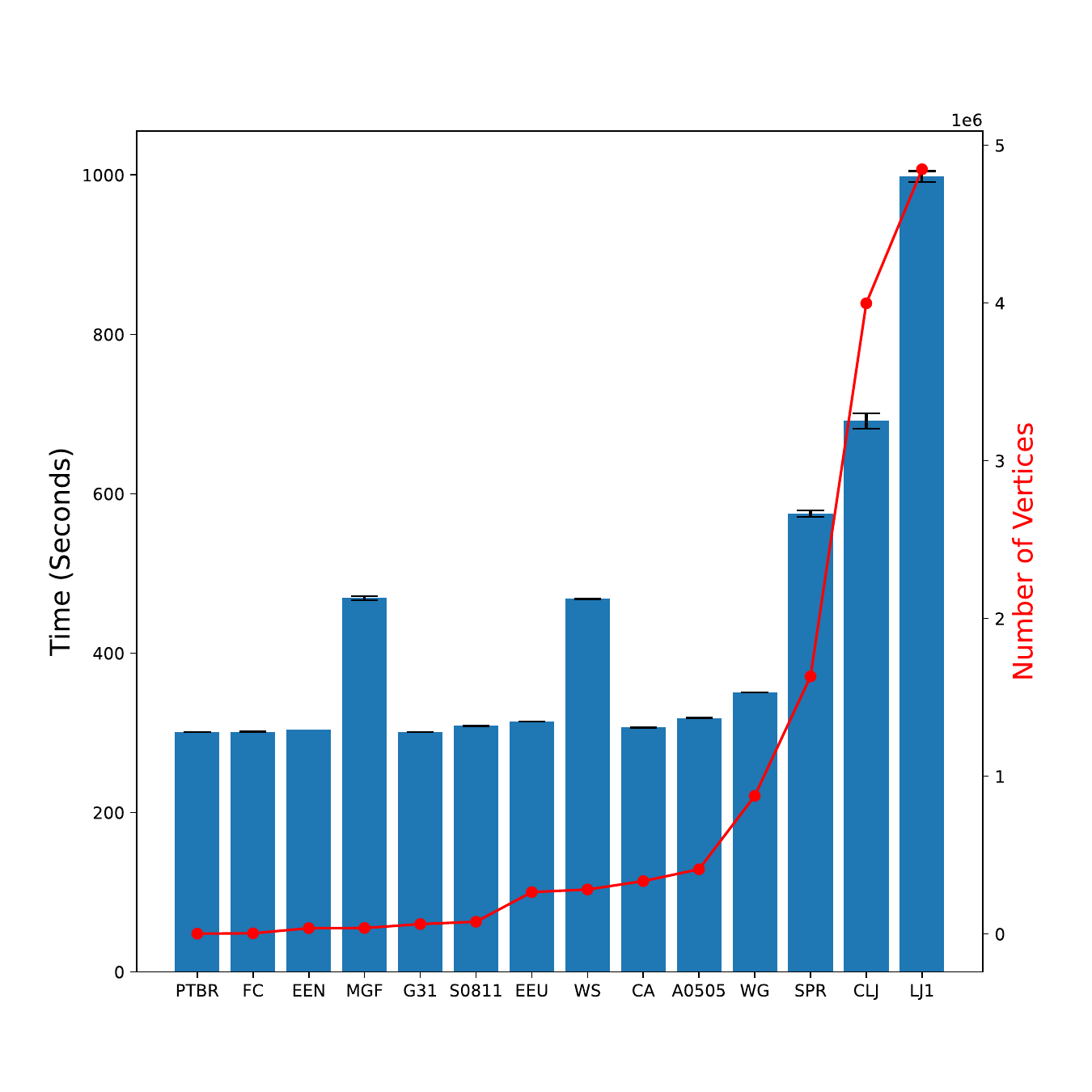}
    \caption{Total Running Time}
    \label{fig:Total Running Time}
\end{figure}

Figure~\ref{fig:Total Running Time} shows the total running time of the experiment for each graph with confidence intervals, where the x-axis is the tested graphs and the y-axis is the total time in minutes for each graph to complete the distributed $k$-core decomposition. Since large graph data requires a long running time, we ran the experiment 20 times for all graph data. The confidence interval varies for each graph data. This experiment is conducted in a single machine instead of a truly distributed environment. All nodes are hosted by the Go routine to simulate the concurrent runtime. Go routines are based on Go scheduling. Large test graphs require a large number of Go routines. For example, graph $LJ1$ requires 4847571 Go routines and there are only 128 threads available on the machine. This increases the workload on the Go scheduler, which causes the variation in the execution time. In addition, the figure has a distribution similar to that in Figure~\ref{fig:Total Number of Passing Messages}, which means that the message passing spends much more time than the computation on the local machine.

However, Figure~\ref{fig:Total Running Time} cannot be used as a reference for real-world implementation. If the distributed $k$-core decomposition algorithm is implemented in the real world, all messages will be sent over the Internet. The geographical locations of the nodes will cause various delays in transferring messages. Our experiment leverages \emph{Go Channels} and computer memory, which generate fewer delays than the Internet. In order to mitigate the problem and make the simulation comparable to the real-world environment, channels with delay can be used. A channel with delay is a channel that sends the message to a copy process that forwards the data to the original destination. The copy process will generate some random or fixed delay to simulate network delays in the real world. In this case, the simulation results can be transferred to the real world.

\subsection{Evaluate the Scalability}
A scalability test is a type of software performance test designed to evaluate the performance of an application, system, or network when its workload increases. We used synthetic graphs with an increasing number of nodes to test the scalability of the algorithm. These synthetic graphs are generated using the \emph{RMAT} (Recursive Matrix) model~\cite{doi:10.1137/1.9781611972740.43}, which mimics the structure of real-world networks, such as social networks, web graphs, and biological networks. It was introduced as part of the Graph 500 benchmark\footnote{\url{https://graph500.org/}} to generate large graphs to test graph algorithms and frameworks. For these generated graphs, the average degree is fixed at 8. 
We use Stanford University's \emph{SNAP} python library~\footnote{\url{https://snap.stanford.edu/snappy/index.html}} to generate RMAT synthetic graphs in Table \ref{rmat graphs} for testing.
\begin{table}
    \centering
    \begin{tabular}{|c|c|c|c|l|l|} \hline 
         Graph&  Node&  Edges& Avg-Degree &Max-Degree&Max-Core\\ \hline 
         RMAT-1000&  1000&  8000&  8& 100&13\\ \hline 
         RMAT-10000&  10000&  80000&  8& 150&17\\ \hline 
         RMAT-100000&  100000&  800000&  8& 399&26\\ \hline 
         RMAT-1000000&  1000000&  8000000&  8& 1132&39\\ \hline 
         RMAT-2000000&  2000000&  16000000&  8& 1378&45\\ \hline
 RMAT-5000000& 5000000& 40000000& 8& 1744&43\\\hline
    \end{tabular}
    \caption{RMAT graphs}
    \label{rmat graphs}
\end{table}

%\subsubsection{Scalability Test Results}
\begin{figure}[!htb]
    \centering
    \includegraphics[width=0.8\linewidth]{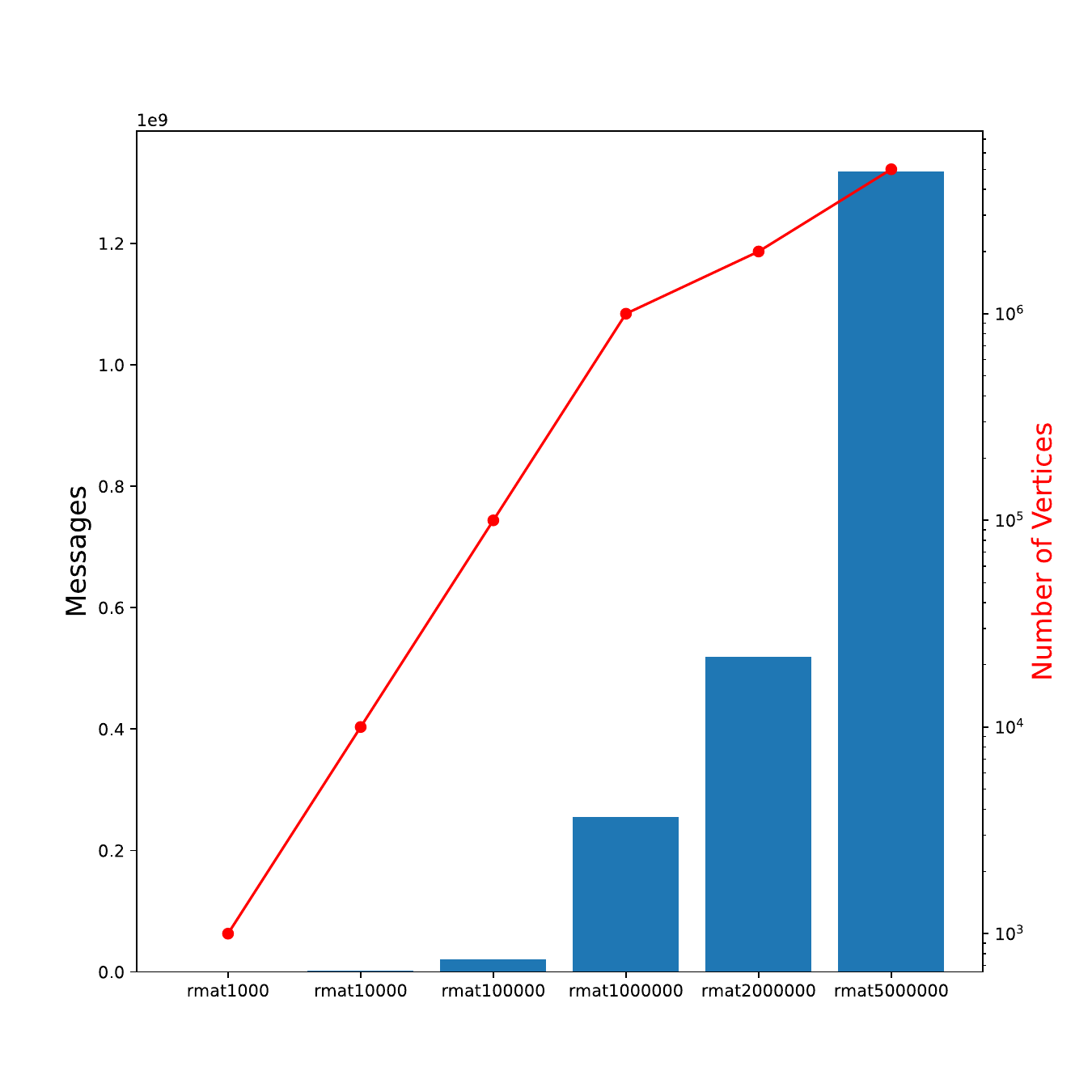}
    \caption{Total Number of Passing Messages }
    \label{fig: scale total message}
\end{figure}

\begin{figure}[!htb]
    \centering
    \includegraphics[width=0.8\linewidth]{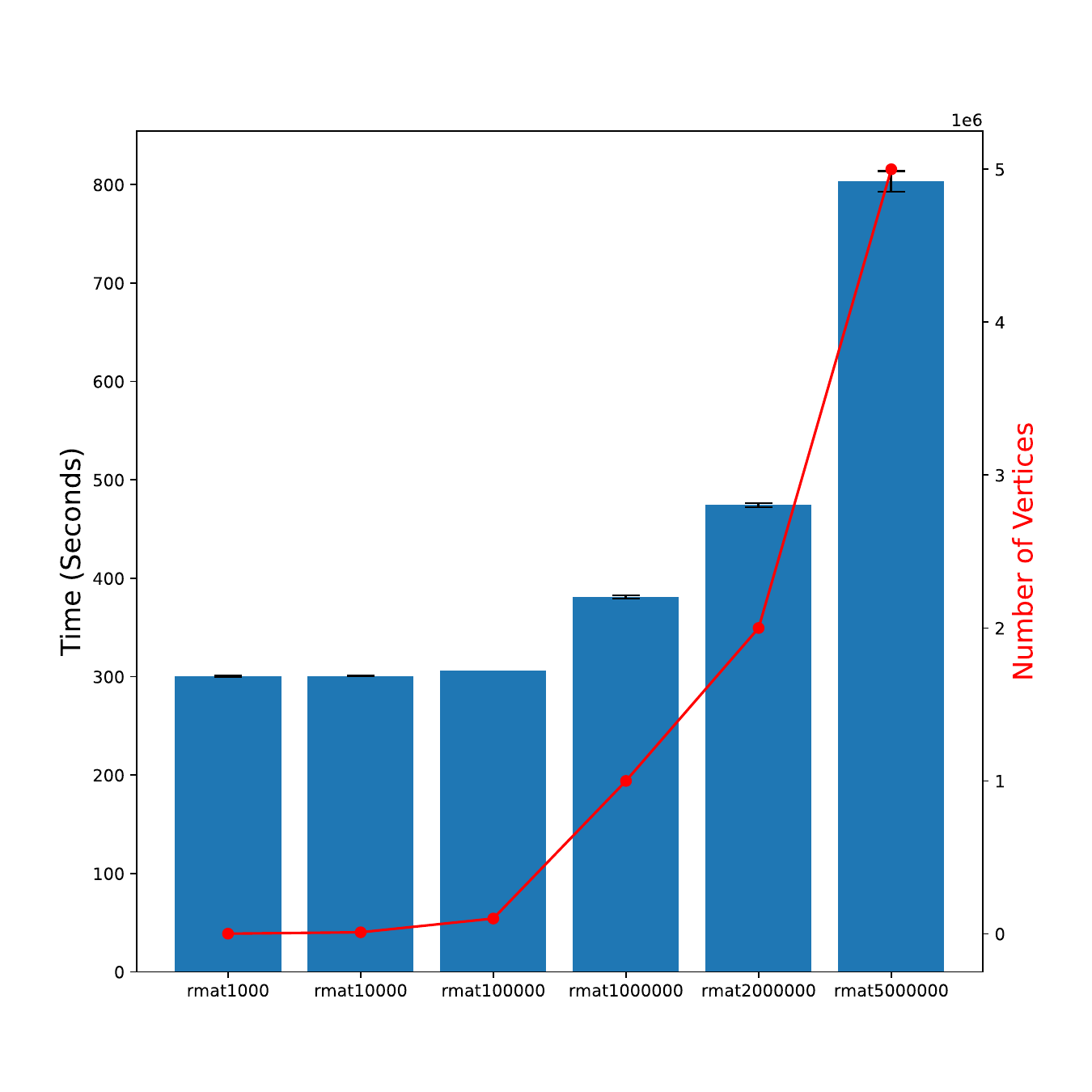}
    \caption{Total Running Time}
    \label{fig: scale total running time}
\end{figure}

Figure~\ref{fig: scale total message} shows the total number of messages passed for each of the synthetic graphs. The number of messages passing increases as the number of nodes increases. This is based on the fact that all synthetic graphs have the same average degree. 
Figure~\ref{fig: scale total running time} shows the total running time of the experiments with synthetic graphs. The confidence interval increases as the number of nodes increases. The performance of the distributed $k$-core decomposition algorithm does not degrade when running on a large scale. 

\iftrue
\subsection{Discussion for Threats to Validity}
\paragraph{External Validity: Sample Characteristics.}
Our experiment uses real-world data graphs. These data graphs are only selected from the SNAP repository. The SNAP dataset collection mainly focuses on large networks. There are other data graphs of different categories in addition to networks.
For example, the results of our experiment may not be generalized to all data graphs in the real world.

For mitigation strategies, we select data graphs from different types of networks to maximize the difference between each data graph. We acknowledge this limitation and suggest that future research should replicate the study in different data graph repositories to verify our findings.

\paragraph{Statistical Conclusion Validity: Low Statistical Power}
Our study has a relatively small sample size due to the time constraint of running the experiment. This could reduce the statistical power, meaning that we may fail to detect a real characteristic of the algorithm.
For example, large graphs require several hours to complete the experiment. We performed the experiment only 20 times for all graphs.

For mitigation strategies, it is recommended that future researchers perform the experiment more than 20 times to expand the results to obtain more accurate observations.
\fi 

\section{Conclusion and Future Work}
\label{conclusion}
This work presents an experimental evaluation of the distributed $k$-core decomposition algorithm on real-world graphs with up to millions of vertices. The algorithm is able to calculate the core number for each node without shared memory. Golang is used to simulate the distributed runtime environment, allowing us to analyze the complexity of the algorithm.
The experiment shows that core decomposition can be done in parallel, even with large data graphs. Although the experiment is conducted on a single machine, it still involves the way messages are passed when running $k$-core decomposition on a real network. We can gain insight into real-world distributed environments, where each vertex represents a client, and there may be millions of clients.  The number of messages passed should be similar between the simulation and the real network. In addition, changes in node status in the experiment reflect how devices in the real network behave.

In the future, our experimental evaluation can be extended to other distributed graph algorithms, such as $k$-truss decomposition and SCC decomposition. Also, instead of using Golang, we can develop a specific framework to simulate distributed algorithms, which supports accurate latency when passing messages.  

\bibliography{reference}
%\printbibliography

\end{document}